\documentclass[10pt,twoside,a4paper]{article}
\usepackage{a4}
\usepackage{graphicx}
\usepackage{amsmath}
\usepackage{amssymb}
\usepackage{latexsym}

\makeindex

\newcounter{mnotecount}[section]

\newcommand{\bref}[1]{(\ref{#1})}
\newcommand{\pd}[2]{\frac{\partial #1}{\partial #2}}
\newcommand{\half}{\tfrac{1}{2}}

\newcommand{\bR}{{\mathbb{R}}}

\newcommand{\cB}{\mathcal{B}}

\newcommand{\DB}{\mathcal{D}_\beta}
\newcommand{\cH}{\mathcal{H}}
\newcommand{\cC}{\mathcal{C}}   
\newcommand{\cF}{\mathcal{F}}   
\newcommand{\cL}{\mathcal{L}}   
\newcommand{\cM}{\mathcal{M}}   

\newcommand{\go}{\mathring{g}}
\newcommand{\gQS}{g}
\newcommand{\divo}{\mathrm{div}_o}
\newcommand{\sdiv}{\mathrm{div}\,}

\newcommand{\II}{I\!I}
\newcommand{\tr}{\mathrm{tr}}
\newcommand{\trM}{\tr_\gM}
\newcommand{\gV}{g{}}           
\newcommand{\gM}{\gamma{}}          
\newcommand{\RiemV}{\mathrm{Riem}^V}  
\newcommand{\RicV}{\mathrm{Ric}^V}    
\newcommand{\RV}{\mathrm{R}^V}    
\newcommand{\RicM}{\mathrm{Ric}^M}    
\newcommand{\RiemM}{\mathrm{Riem}^M}  
\newcommand{\RM}{\mathrm{R}^M}    
\newcommand{\extK}{K}           
\newcommand{\trK}{\trM\extK}    
\newcommand{\gradV}{D{}}   
\newcommand{\gradM}{\nabla}   
\newcommand{\gradN}{\hat{\nabla}}   
\newcommand{\Hess}{\mathrm{Hess}}   
\newcommand{\al}{a}  
\newcommand{\be}{b}  
\newcommand{\ga}{c}  
\newcommand{\de}{d}  
\newcommand{\ep}{e}  

\title{\bf  The Constraint Equations}

\author{Robert Bartnik\thanks{%
School of Mathematics and Statistics,
University of Canberra,
ACT 2601, Australia.  E-mail: robert.bartnik@canberra.edu.au}
\\
Jim Isenberg\thanks{%
Department of Mathematics and Institute for Theoretical Science,
University of Oregon,
Eugene, OR 97403, USA. E-mail: jim@newton.uoregon.edu}
}

\date{19th December 2003}

\begin{document}
\maketitle

\begin{abstract}
Initial data for solutions of Einstein's gravitational field equations  
cannot be chosen freely: the data must satisfy the four Einstein  
constraint equations. We first discuss the geometric origins of the  
Einstein constraints and the role the constraint equations play in  
generating solutions of the full system. We then discuss various ways  
of obtaining solutions of the Einstein constraint equations, and the  
nature of the space of solutions.

\vskip 4.5mm



\end{abstract}

\vskip 12mm

\section{Introduction}
Yvonne Choquet-Bruhat's epic work of over 50 years ago shows that if a
set of smooth initial data which satisfies the Einstein constraint
equations is given, then we can always find a spacetime solution of
the Einstein equations which contains an embedded hypersurface whose
metric and second fundamental form agree with the chosen data. In the
years since then, arguably the most important method for constructing
and studying solutions of Einstein's equations has been the initial
value (or Cauchy) formulation of the theory, which is based on
Yvonne's result. Especially now, with intense efforts underway to
model astrophysical events which produce detectable gravitational
radiation, the Cauchy formulation and numerical efforts to implement
it are of major interest to gravitational physicists.

To understand the Cauchy formulation of Einstein's theory of  
gravitation, we need to understand the constraint equations. Not only  
do the constraints restrict the allowable choices of initial data for  
solutions; they also effectively determine the function space of  
maximally globally hyperbolic solutions of the theory, and they play a  
role in generating the evolution of the initial data via their  
appearance in the Hamiltonian for Einstein's theory.

The goal of this review paper is to provide some measure of
understanding of the Einstein constraints. We start in Section 2 by
explaining the geometric origin of the constraint equations. To do
this, we discuss $3+1$ foliations of spacetimes, the Gauss-Codazzi
decompositions of the curvature, and the consequent $3+1$ projection
of the spacetime Einstein equations. Next, in Section 3, we discuss
the relationship between the constraints and evolution. Here, after
first reviewing the proof of well-posedness, we discuss the ADM
Hamiltonian formulation of Einstein's theory, noting the relationship
between the Hamiltonian functional and the constraints. We also
examine the evolution equations for the constraints, noting that if a
set of data satisfies the constraints initially, it will continue to
satisfy them for as long as the evolution continues.

We discuss methods for constructing solutions of the Einstein
constraint equations in Section 4. We focus first on the most useful
approach to date: the conformal method. After describing how the
conformal method works, we discuss its success in parameterizing the
set of all constant mean curvature (CMC) solutions of the constraints,
the difficulties which arise when constructing solutions with
non-constant mean curvature, and the issue of finding physically
relevant sets of data which satisfy the constraints.

Closely related to the conformal method is the conformal thin sandwich
approach.  We describe how it relates to the standard conformal
method, and  its major advantages and disadvantages.


Instead of exploiting conformal variations to enforce the constraints,
the original thin sandwich formulation
\cite{BaierleinEt62,BelascoOhanian69} varies the lapse (via an
algebraic relation) and the shift vector.  Under certain conditions
this procedure leads to an elliptic system for the shift
\cite{BartnikFodor93}, and an implicit function theorem then shows
that the system is solvable for all nearby data.  This constructs an
open set of solutions of the constraint equations, from unconstrained
data.  However, the restrictions arising from the surprising
ellipticity condition, that $\pi^{ij}$ be positive or negative
definite, mean that the original thin sandwich approach is viable only
for a limited range of geometries, and at this stage it must be
considered more of a curiosity than a practical solution technique.


It is also possible to construct solutions of the Hamiltonian
constraint by solving a semi-linear parabolic equation.  Recent work
\cite{ShiTam02,SmithWeinstein00,SmithWeinstein03} has shown that the
essential feature of the original ``quasi-spherical'' construction
\cite{Bartnik93} can be generalized beyond the quasi-spherical
foliation condition to give a flexible technique for constructing
exterior metrics of prescribed scalar curvature, satisfying geometric
inner boundary conditions.  Whilst the resulting parabolic method is
restricted by the requirement that the background must be
quasi-convex, it is able to handle several questions involving the
constraints which cannot be addressed using the conformal method.

We finish our review of constraint construction techniques by
discussing the idea of gluing, and how this idea has been implemented
to date. Two approaches to gluing have been developed and applied to
solutions of the constraint equations. Connected sum or IMP gluing
\cite{IsenbergMazzeoPollack02} builds new solutions by adding a
cylindrical bridge (or wormhole) connecting a pair of points either on
a single given solution or on a pair of given solutions.  The
Corvino-Schoen technique
\cite{Corvino00,CorvinoSchoen03,ChruscielDelay03} gives a local
projection from approximate solutions to exact solutions of the
constraints.  This may then be applied to smoothly attach a finite
region in a given asymptotically Euclidean solution to an exterior
Schwarzschild or Kerr solution, with a transition region connecting
the two regions. Both approaches have been very useful in answering a
number of longstanding questions regarding solutions of the
constraints.

We conclude this paper with a number of comments on important
issues concerning the constraints which need to be addressed.

\section{Deriving the Constraints}
\label{sec:geometric-background}

The Cauchy formulation is used primarily to construct new solutions of
the Einstein gravitational field equations from specified initial
data.  The best way to understand the origins of the constraint
equations is to assume that we have a spacetime solution and to consider
the induced data on spacelike hypersurfaces.  We do this here, showing
that the constraint equations necessarily must  be satisfied by initial
data $(M^3, \gM, \extK)$, where $M^3$ is a manifold, $\gM$ is a
Riemannian metric on $M^3$ and $\extK$ is a symmetric tensor on $M^3$,
if this data is to be induced by a spacelike hypersurface in a
spacetime solution of Einstein's equations.

\subsection{ $3+1$ Foliations of Spacetimes}

Let $V^4$ be a smooth 4-dimensional Lorentzian spacetime, with the
smooth\footnote{Although it is useful in addressing certain questions
  to consider lower regularity, for our present purposes it is most
  convenient to assume that the spacetime and the metric are
  $C^{\infty}$.} metric $\gV$ having signature $(-1,1,1,1)$. A
\emph{hypersurface} in $V$ is an (embedded) submanifold $M^3
\hookrightarrow V^4$ of codimension 1.  $M$ is \emph{spacelike} if the
induced bilinear form $\gM := i^*\gV$ is a Riemannian metric on $M$.
Equivalently, $M$ is spacelike if at each point $x\in M$ there is a
timelike future unit normal vector $n$.  We have the familiar diagram:
\begin{center}
\resizebox{0.7\textwidth}{!}{\includegraphics{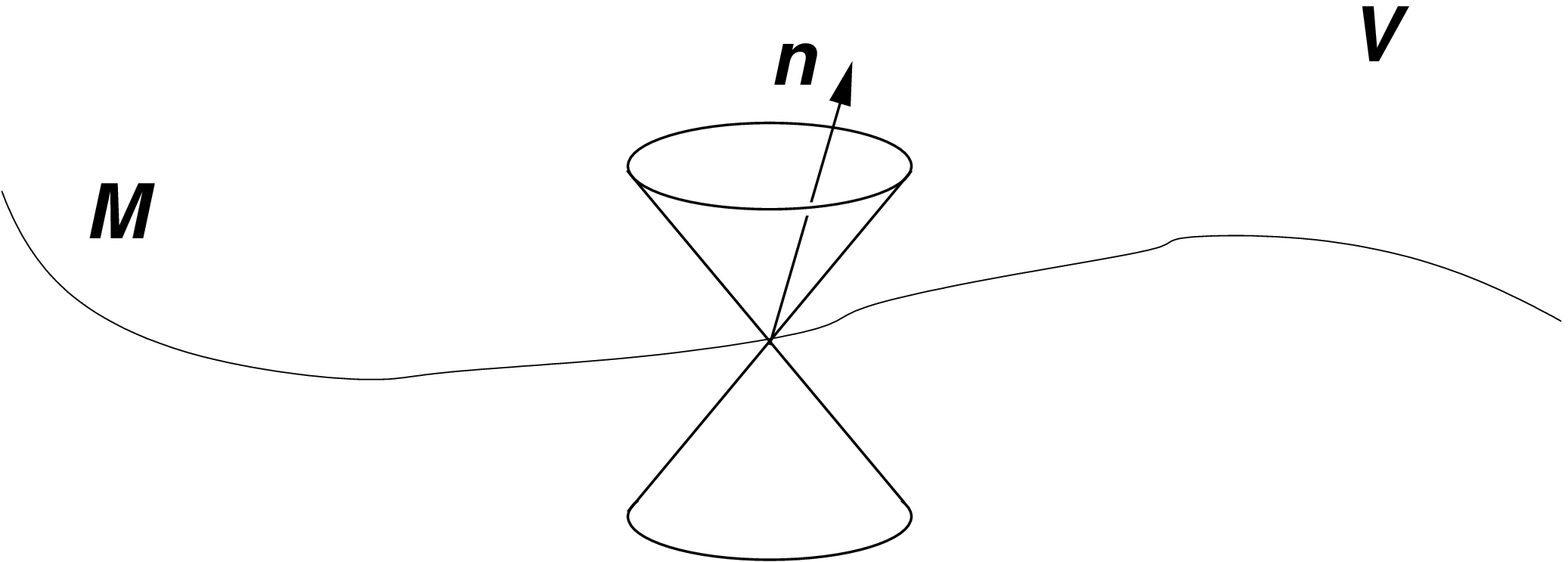}}
\end{center}
If $X,Y$ are any vector fields tangent to $M$, then as a consequence
of the embedding, we can consider them as vectors in $V$ and we can
decompose the $V$-covariant derivative $\gradV_XY$ into tangential and
normal components,
\begin{equation}
    \gradV_XY = \gradM_XY +\extK(X,Y) n,
\end{equation}
where $\gradM$ is the induced connection on $M$ (it is also the
Levi-Civita connection of the induced Riemannian metric on $M$), and
$\extK$ is a bilinear form (rank 2 tensor) on $M$, called the
\emph{second fundamental form}, or \emph{extrinsic
  curvature}\index{extrinsic curvature} in the physics literature.
From the relation
\begin{equation}
    \extK(X,Y) = \gV(\gradV_Xn,Y)
\end{equation}
and from the fact that the Lie bracket of $X$ and $Y$ is tangent to $M$,
$[X,Y]\in TM$, we find that $\extK$ is a symmetric form,  
$\extK(X,Y)=\extK(Y,X)$.

A function $t\in C^1(V)$ is a \emph{time function}\index{time
  function} if its gradient is everywhere timelike.  We say that a
time function $t$ is \emph{adapted to the hypersurface} $M$ if $M$ is
a level set of $t$, in which case we can choose adapted local
coordinates $(x,t)$.  In terms of such coordinates the normal vector
field takes the familiar lapse-shift\index{lapse}\index{shift vector}
form
\begin{equation}
    n = N^{-1}(\partial_t - X^i\partial_i)\;,
\end{equation}
where $N$ is the \emph{lapse function} and $X = X^i\partial_i$ is the
\emph{shift vector}. Equivalently,
\begin{equation}
    \partial_t = Nn+X.
    \end{equation}
This time evolution vector field $\partial_t$ need not necessarily be  
timelike everywhere. (The shift vector $X\in TM$ is of course  
necessarily spacelike wherever it is nonvanishing.) However, if
$\partial_t$ is timelike   
then the $x=const.$~paths together make up a timelike congruence of  
spacetime observers.
\begin{center}
   \resizebox{0.7\textwidth}{!}{\includegraphics{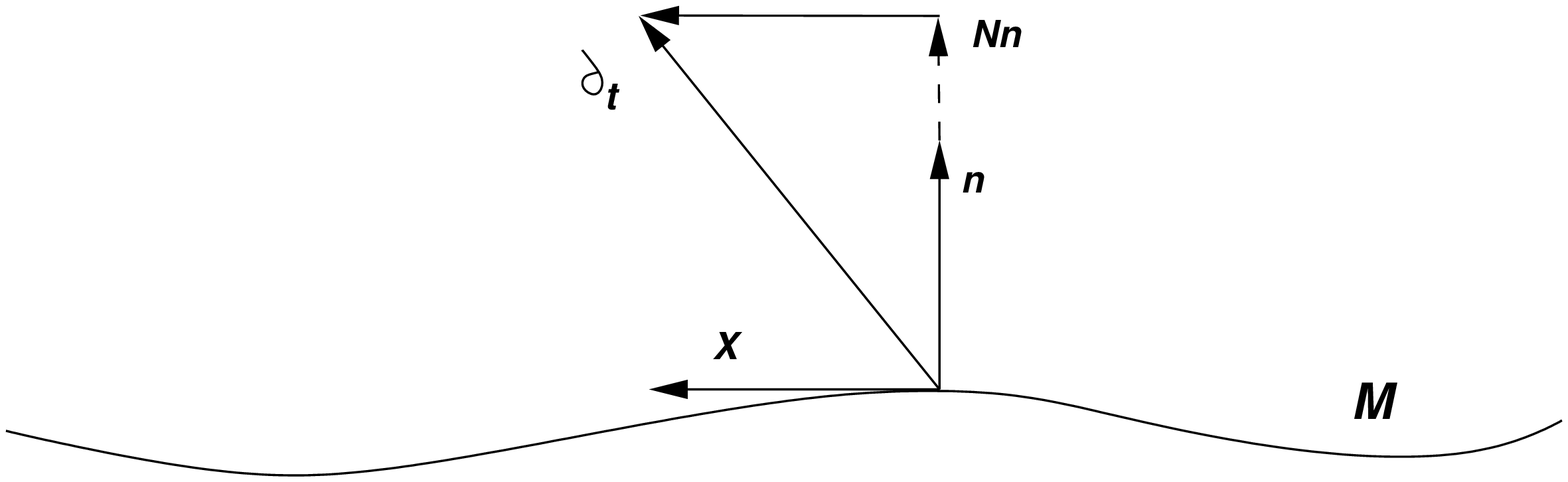}}
\end{center}
The spacetime metric can be expressed in terms of the lapse and the
shift and the spatial metric in adapted coordinates by the
formula\footnote{Our index conventions here are: mid latin alphabet
  indices ($i, j$, etc.) run from 1 to 3 and correspond to directions
  tangent to the spacelike manifold $M$, while early latin indices ($a,
  b$, etc.) run from 0 to 3 and correspond to spacetime directions.}
\begin{equation}
   \label{eq:gADM}
    \gV = -N^2\,dt^2 + \gM_{ij}(dx^i + X^i\,dt)(dx^j + X^j\,dt);
\end{equation}
and the second fundamental form is given by
\begin{equation}
    \extK_{ij} = \extK(\partial_i,\partial_j)
    = \half N^{-1}(\partial_t \gM_{ij}- \cL_X\gM_{ij})
\end{equation}
where
\begin{equation}
    \cL_X\gM_{ij} = \gradM_iX_j + \gradM_jX_i
\end{equation}
is the Lie derivative in $M$ of the spatial metric $\gM$ in the  
direction
$X\in TM$. Thus we obtain an expression for the $\partial_t$ evolution  
of
the spatial metric,
\begin{equation} \label{dtg}
    \partial_t\gM_{ij} = 2N\extK_{ij}+\cL_X\gM_{ij}.
\end{equation}

\subsection{The Gauss and Codazzi Equations and the Constraints}

Based on the orthogonal decomposition of the covariant derivative
operator $\gradV=\gradM +\extK$, and on the definition of the
curvature on $V$, the Gauss and Codazzi equations relate certain
spacetime curvature components to spatial curvature and other
expressions formed solely from data intrinsic to the submanifold $M$.
For example, starting from the expression for the curvature on $V$
\begin{equation}
    \RiemV(X,Y,Z,W) =
    \gV((\gradV_X\gradV_Y-\gradV_Y\gradV_X-\gradV_{[X,Y]})Z,W) ,
\end{equation}
if  $X,Y,Z,W$ are all  vector fields tangent to $M$, then we find by a  
simple
computation that
\begin{eqnarray}
\nonumber
\lefteqn{    \RiemV(X,Y,Z,W) \hspace*{3cm} }
\\
\label{gauss}
 &=&  \RiemM(X,Y,Z,W)  
+\extK(X,W)\extK(Y,Z)-\extK(X,Z)\extK(Y,W).
\end{eqnarray}
\index{Gauss equation}
This is the \emph{Gauss equation}; it shows that the intrinsic
curvature of $M$, measured by $\RiemM$, is determined by curvature
components of the ambient spacetime $V$ and the second fundamental
form $\extK$.  In the perhaps more familiar index notation, the Gauss
equation takes the form
\begin{equation}
    \RV_{ijkl} = \RM_{ijkl} +\extK_{il}\extK_{jk}-\extK_{ik}\extK_{jl},
\end{equation}
where the indices $i,j,k,l$ refer to an (unspecified) basis of the spatial
tangent space $TM$, which may be equally well determined either by a
coordinate basis $\partial_i$ (holonomic basis) or by an orthonormal
basis $e_i$, $i=1,\dots,3$.

A similar computation gives the \emph{Codazzi identity},\index{Codazzi
  equation}
\begin{eqnarray}
   \RiemV(X,Y,n,Z) &=& \gV((\gradV_X\gradV_Y - \gradV_Y\gradV_X
   -\gradV_{[X,Y]}) n,Z)
\nonumber
\\
\label{codazzi}
&=& \gradM_X\extK(Y,Z) - \gradM_Y\extK(X,Z)\;,
\end{eqnarray}
where
\begin{equation}
    \gradM_X\extK (Y,Z) = D_X(\extK(Y,Z)) -  
\extK(\gradM_XY,Z)-\extK(Y,\gradM_XZ)
\end{equation}
is the covariant derivative in $M$ of the tensor $\extK$.  In index
notation the Codazzi identity takes the form
\[
\RV_{ijnk} = \extK_{jk;i} - \extK_{ik;j}\;.
\]
What about  $\RiemV(n,Y,n,Z)$?  A calculation similar to those used to  
derive the Gauss and Codazzi equations does not lead to an expression  
solely in terms of $\gM$, $\extK$, and their spatial derivatives. If
we introduce the Lie derivative  
\[
\cL_n\extK(Y,Z) = D_n(\extK(Y,Z)) - \extK(\cL_nY,Z) - \extK(Y,\cL_nZ),
\]
along with the lapse and its spatial derivatives, then we have
\index{Mainardi equation}
\begin{equation}
   \RiemV(Y,n,n,Z) = -\cL_n\extK(Y,Z) + \extK^2(Y,Z)+N^{-1}  \Hess(N)(Y,Z),
\label{mainardi}
\end{equation}
where $\extK^2_{ij} := \extK_i^k\extK_{jk}$ and $\Hess(N)=\nabla^2N$
is the Hessian or second covariant derivative of the lapse function
$N$.  This is sometimes referred to as the \emph{Mainardi equation},
and has the index form 
\begin{equation} \label{mainardi2}
\RV_{innj} = -\cL_n\extK_{ij} +\extK_i^k\extK_{jk}+ N^{-1}N_{;ij}.
\end{equation}
The index expressions for the Gauss, Codazzi and Mainardi equations
given above are useful for calculating the $3+1$ expressions for
contractions of the Riemann tensor, such as the Ricci curvature
tensor, the scalar curvature, and the Einstein tensor. For example,
the Einstein tensor $ G_{\al\be} = \RicV_{\al\be} -\half \RV
\gV_{\al\be} $ satisfies
\begin{eqnarray}\label{C1}
    2G_{nn} &=&  \RM + (\trM  \extK)^2 - ||\extK||^2 \\
     G_{in} &=&  \extK_{i;j}^{j}-\extK^j_{j;i}.
\end{eqnarray}
We note that these expressions for $G_{nn}$ and $G_{in}$ involve just
$\gM, \extK$, and their spatial derivatives. Hence it follows from the
Einstein field equations
\begin{equation}\label{ee}
G_{ab}=8\pi T_{ab}
\end{equation}
that the $G_{nn}$ and $G_{in}$ equations are constraints on the choice
of the initial data $\gM$ and $\extK$.  These are the \emph{Einstein
  constraint equations}\index{constraint equations}
\begin{eqnarray}\label{Hcon}
     \RM + (\trM  \extK)^2 - ||\extK||^2 &=& 16\pi T_{nn},
\\
\label{pcon}
    \extK_{i;j}^{j}-\extK^j_{j;i} &=& 8\pi T_{ni}.
\end{eqnarray}
Here $T_{ab}$ is the stress-energy tensor and describes the matter
content of the ambient spacetime.  Although $T_{ab}$ will itself
generally satisfy some evolution equation, we will regard it as a
prescribed field and often focus for definiteness on the case
$T_{ab}=0$ of vacuum (matter-free) spacetimes.

It is clear from our discussion here that the constraint equations are
conditions on the data $(M,\gM,\extK)$ which are \emph{necessary} for
the data to arise from a spacelike hypersurface in a spacetime
satisfying the Einstein field equations (\ref{ee}).  That these
conditions are also \emph{sufficient} (for the existence of a
spacetime satisfying the Einstein equations and which induces the data
on some hypersurface) is the content of the fundamental theorem of
Y.~Choquet-Bruhat \cite{FouresBruhat52}, which we discuss in the next
section.

\section{The Constraints and Evolution}
The constraint equations comprise four of the ten Einstein equations.  
The remaining six equations
\begin{equation}
\label{evolution}
G_{jk}=8\pi T_{jk}
\end{equation}
describe how the data $(\gM,\extK)$ evolve in a spacetime solution.
This can be seen by rewriting the curvature terms in \bref{mainardi},
giving 
\begin{equation}
  \label{Kevolution}
  \cL_n\extK_{ij} = G_{ij}-\half G_c^c \gV_{ij} +2\extK^2_{ij} -
  \trM\extK \extK_{ij}-\RicM_{ij} +N^{-1}N_{;ij}.
\end{equation}
Yet, as noted in the introduction, the constraints play a role in the
evolution as well.  We discuss this role here. We first show that the
constraints are sufficient as well as necessary for a spacetime
solution to evolve from a given set of data. We next explore the role
of the constraints in the canonical, Hamiltonian, formulation of
Einstein's theory. Finally, we discuss the evolution of the constraint
functions under the evolution of the data generated by
(\ref{evolution}).

\subsection{Well-posedness of the Vacuum Einstein Equations}

The Cauchy formulation of a given field theory is well-posed if (a)
for any choice of initial data of specified regularity, there exists a
solution which is consistent with that data, and (b) the map from the
space of initial data to solutions is continuous. As shown by
Choquet-Bruhat \cite{FouresBruhat52}, the Cauchy formulation of
Einstein's (vacuum) theory of gravitation, for smooth initial data, is
well-posed in coordinates of wave map gauge (harmonic) type. We now
review the proof of this result
\cite{Lanczos22,FouresBruhat52,ChoquetYork79,DeTurck83,Friedrich85,Taylor96}.

The vacuum Einstein equations take the form $\RicV(\gV)=0$. If this
were a hyperbolic system, well-posedness would follow from general
results on quasi-linear hyperbolic systems (eg.~\cite{John82,Taylor96}).
However, the system is not hyperbolic, which we verify by noting
that by  diffeomorphism invariance we have $\Phi^*(\RicV(\gV)) =
\RicV(\Phi^*(\gV))$ for any diffeomorphism $\Phi$ of $V$, so the
symbol (leading order terms in the linearisation) of $\RicV(\gV)$,
considered as a partial differential equation in local coordinates,
has a very large kernel.

To get around this difficulty, we consider the \emph{reduced Einstein
  equations} \cite{Lanczos22,deDonder26}\index{Einstein equations!reduced}
\begin{equation}\label{rho}
    \rho(\gV) := \RicV(\gV) + \half\cL_W \gV =0,
\end{equation}
where $\cL_W\gV_{\al\be} = \gradV_\al W_\be +\gradV_\be
W_\al$ is the Lie derivative of the metric $\gV$ in the direction of
the vector field $W$, and $W$ is chosen to have leading terms
\begin{equation}
   \label{W}
    W_\al =  - \gV^{\ga\de}(\partial_\ga \gV_{\de\al}
    - \half \partial_\al \gV_{\ga\de}) + F(\gV)
\end{equation}
where the  terms $F(\gV)$  do not involve derivatives of $\gV$.  A short
calculation shows that $\rho(\gV)$ has symbol $\half
\gV^{\al\be}\xi_\al\xi_\be$, exactly that of the wave equation of the
Lorentzian metric $\gV$ itself.  Thus $\rho(\gV)=0$ forms a
quasi-linear hyperbolic system,
\begin{equation}\label{qlhyp}
    \gV^{\ga\de}\partial^2_{\ga\de}\gV_{\al\be} =
    Q_{\al\be}(\gV,\partial \gV),
\end{equation}
where $Q_{\al\be}$ is quadratic in $\partial \gV$ and depends also on
the $F(\gV)$ terms in (\ref{W}).  As noted above, the initial value problem
for $\rho(\gV)=0$ is well-posed.

However in general solutions of $\rho(\gV)=0$ will not satisfy the
vacuum Einstein equations.  The idea now is to choose the initial
conditions (and, perhaps, the precise form of the terms $F(\gV)$) in
such a way that the vector field $W$ can be shown to vanish
identically.

To understand the meaning of $W$, recall the definition of the 
\emph{tension field} $\omega(\phi)$ of a map $\phi:(V,\gV)\to
(N,\gradN)$,
\begin{equation}
   \label{tau}
   \omega^\alpha(\phi) =   \gradN^a D_a\phi^\alpha =
   \gV^{\ga\de}\left(\partial_{\ga\de}^2\phi^\alpha - \Gamma_{\ga\de}^\ep
     \partial_\ep\phi^\alpha  + \hat{\Gamma}_{\gamma\delta}^\alpha
     \partial_\ga\phi^\gamma \partial_\de \phi^\delta\right) ,
\end{equation}
where $N$ is some chosen four dimensional manifold and $\gradN$ is an
arbitrary connection on $N$.  If we now assume $\phi$ is a
diffeomorphism, which we use to identify the coordinates $x^{\al}$ on
$V$ with $y^\alpha$ on $N$ by pullback $x^\al =\phi^*(y^\al)$, then
the tension becomes
\begin{equation}
\omega^\al(\phi) = - \gV^{\al\be}\gV^{\ga\de}(\gradN_\ga
    \gV_{\de\be} - \half \gradN_\be \gV_{\ga\de})\;,
\end{equation}
which shows by comparison with \bref{W} that $W$ may be chosen as the
metric dual of the push-forward by $\phi^{-1}$ of the tension,
\begin{equation}
   \label{eq:17}
   W_{\al} = \gV_{\al\be} \phi^{-1}_*(\omega^\be(\phi))\;.
\end{equation}
Note that the reduction to a quasi-linear hyperbolic system holds for
any choice of the target $(N,\gradN)$, where the connection $\gradN$
need not be compatible with any metric on $N$.  In the simplest case
$\gradN$ is the flat connection on $\bR^4$ and the condition $W=0$
translates into $\Box_\gV x^\al =0$ in local coordinates $x^\al$ on
$V$ and $N$.  The gauge condition $W(\phi)=0$ is known as the
\emph{wave map} or \emph{Lorentz-harmonic} coordinate gauge\index{wave
  map gauge}.  The common terminology ``harmonic gauge'' is
inappropriate since solutions of the wave equation do not have the
mean-value (``harmonic'') property implied by the corresponding
Euclidean Laplace equation.

Good existence theorems for the Cauchy initial value problem for the
reduced Einstein system \bref{qlhyp} are well-known (see, for example,
\cite{Taylor96,John82,Racke92}), and the initial data
$(\gV_{\al\be}(0),\partial_t\gV_{\al\be}(0))$ are freely prescribable,
subject only to the condition that $\gV_{\al\be}(0)$ has Lorentz
signature with $M_0$ spacelike.  In particular, if the geometric data
$(\gM_{ij},\extK_{ij})$ are given (subject only to the vacuum
constraint equations (\ref{Hcon})-\ref{pcon})), then for any choice of
the lapse and shift $\gV_{0\al}$ we may recover $\partial_t \gV_{ij}$
via \bref{dtg}.  Since $\gV_{ij}=\gM_{ij}$, we see that initial data
$(\gV_{\al\be}(0),\partial_t\gV_{\al\be}(0))$ may be chosen consistent
with $(\gM,\extK)$ for any choice of
$(\gV_{0\al}(0),\partial_t\gV_{0\al}(0))$.

Of course, it is not sufficient to just solve the reduced Einstein
equations; in order to recover a solution of the full vacuum Einstein
equations we must also ensure that the solution satisfies $W=0$.
However, it is an easy consequence of the second Bianchi identities
and $\rho(\gV)=0$ that $W$ must satisfy
\begin{equation}\label{boxtau}
    \Box_g W + \RicV(\gV)W =0.
\end{equation}
Thus we aim to construct the full initial data
$(\gV_{\al\be}(0),\partial_t\gV_{\al\be}(0))$ for the evolution
equation $\rho(\gV)=0$ in such a way that $W(0)=0$ and $\partial_t
W(0)=0$.  If we can do this, then uniqueness for solutions of the
Cauchy problem for the hyperbolic evolution equation \bref{boxtau}
will imply that $W=0$ everywhere.

To ensure that $W(0)=0$ we may either  choose the background metric
$h$ on $N$ appropriately or we may expand \bref{W} and choose the
components $\partial_t \gV_{0\al}(0)$ appropriately.  As noted above,
the choice of $\partial_t \gV_{0\al}(0)$ does not affect the
prescribed geometric data $(\gM,\extK)$.

To find the conditions necessary for choosing initial data with
$\partial_tW(0)$ we combine the reduced Einstein equation \bref{rho}
and the condition $W(0)=0$ to obtain $-2\RicV_{\al\be}=\partial_\al
W_\be+\partial_\be W_\al$ at $t=0$, which may be rearranged into the
form
\begin{equation}
\label{Wdot}
g^{00}\partial_t W_\al = -2 G^0_\al +
(\mathrm{ terms\  vanishing\  at\ }\ t=0).
\end{equation}
Then, noting that $G^0_\al $ are the constraint functions, we see from
\bref{Wdot} that the condition that the geometric data $(\gM,\extK)$
satisfy the constraint equations ensures that $\partial_t W(0)=0$.
Since solutions to \bref{boxtau} are uniquely determined by the
initial data $(W(0), \partial_t W(0))$, the corresponding solution of
the reduced equation \bref{rho} will have $W=0$ everywhere and thus
$\gV$ will satisfy the vacuum Einstein equations with geometric data
$(\gM,\extK)$.  This shows that the constraint equations are also
sufficient for the existence of a compatible vacuum spacetime, as
claimed.


In this formulation of the Einstein equations, the lapse and shift are
determined from $g_{0a}(t)$ by solving the reduced Einstein equations.
Alternatively, we may ask whether it is possible to \emph{prescribe}
the lapse and shift as given functions of $x$ and $t$ (for example,
$N(x,t)=1$ and $X(x,t)=0$ for all $x\in M$ and $t\ge0$).  As shown
recently by Choquet-Bruhat \cite{CB01}, this is possible although in a
weaker sense than the Einstein system in the wave map gauge
described above. Specifically, the Einstein system for general lapse
and shift is ``non-strictly hyperbolic''; as a consequence, the system
is well-posed in Gevrey spaces, and so has local existence. We note as
well that a number of researchers \cite{ChoquetYork79,Friedrich85}
have formulated extended first order versions of Einstein's vacuum
equations and have verified symmetric hyperbolicity and therefore
well-posedness in more standard spaces for these versions.

Because the initial lapse-shift variables $g_{0a}(0)$
(cf.~\bref{eq:gADM}) are freely specifiable in the above construction,
there remains the question of the uniqueness of the spacetime
development of a given set of initial data.  It turns out that
solutions constructed using different choices of the lapse and the
shift but the same geometric data $(\gM,\extK)$ are in fact equal in
some neighbourhood of the initial surface, after suitable coordinate
changes \cite{ChoquetBruhatGeroch69}.  This property is established by
constructing a wave map $\psi$ from one solution to the other, and
using the identification of the initial data to ensure the wave map
initially at least is a spacetime isometry.  The two metrics can then
be compared directly (via $\gV_1$ and $\psi^*(\gV_2)$), and uniqueness
for the reduced Einstein equations leads to geometric uniqueness.
This shows that the geometric data $(\gM,\extK)$ determine the
resulting spacetime uniquely, at least locally.


\subsection{Einstein's Theory as a Dynamical System and the ADM  
Hamiltonian Formulation}
\label{sec:adm-hamiltonian}

If we fix a choice of the lapse and shift for which the system of
Einstein's equations is well-posed (see the discussion above), then
Einstein's equations can be viewed formally as a dynamical system.
More specifically, fixing a three dimensional manifold $M$ and letting
$T\cM$ denote the tangent space to the space $\cM$ of smooth
Riemannian metrics on $M$, we find that (with a fixed choice of lapse
$N$ and shift $X$) the Einstein evolution equations
\begin{eqnarray}
   \label{eq:dtga}
   \pd{}{t}\gM_{ij} &=& 2N\extK_{ij}+\cL_X\gM_{ij}\;,
   \\
   \label{eq:dtK}
   \pd{}{t}\extK_{ij} &=&\gradM{}^2_{ij}N + \cL_XK_{ij} 
   + N(\RicV_{ij} - \RicM_{ij}
   + 2\extK_i^k\extK_{jk} - \trM \extK \,\extK_{ij} )
\end{eqnarray}
specify a flow on $T\cM$. The $\extK_{ij}$ evolution equation follows
from the Mainardi equation, cf.~\bref{mainardi} and \bref{Kevolution}.
For the vacuum Einstein case $\RicV_{ij}=0$; in the nonvacuum case,
$\RicV_{ij}$ depends on the matter fields through the stress-energy
tensor $T_{ab}$.

There is an important sense in which the constraints generate the
flow, which we see by recasting Einstein's theory in the canonical
Hamiltonian form. Originally motivated by attempts to obtain a theory
of gravity consistent with quantum theory, and often labeled the
``ADM'' Hamiltonian formulation because of the early work on these
ideas by Arnowitt, Deser and Misner \cite{ADM62}, this  is
based on the analysis of the Einstein-Hilbert Lagrangian $\int_V
R^V\,dv_\gV$.

One way to approach this analysis is via the second variation formula
for spacelike hypersurfaces \cite{Simon83,Bartnik84}; for more
traditional treatments see \cite{ChoquetYork79,Wald84,WaldIyer94}.  We
let $F:(-\epsilon,\epsilon)\times M\to V$ be a 1-parameter family of
embeddings $s:M\mapsto M_s=F(s,M)$, with variation vector
$Y=F_*(\pd{}{s})$, with normal vector $n=n_s$ and with second
fundamental form $\extK=\extK_s$ along $M_s$.  If $Y$ is everywhere
transverse to $M$ then $s\mapsto M_s$ is a foliation and we can regard
$Y, n$, and $\extK$ as fields on $V$; otherwise they live naturally on
$(-\epsilon,\epsilon)\times M$ via pullback.  The first variation of
the area $|M_s|$ is given by
\begin{equation}
   \label{eq:19}
   D_Y(dv_\gM) = - \gV(Y,n)\trM \extK\ dv_\gM\;,
\end{equation}
and the second variation formula is\index{second variation|Lorentzian}
\begin{equation}
   \label{eq:2var}
   D_Y \trM  \extK = \gV(Y,n)(\|\extK\|^2+\RicV(n,n)) +
   \gV(Y,\gradM \trM \extK) - \Delta_M \gV(Y,n)\;,
\end{equation}
where $D_Y=\pd{}{s}$ and the $s$-dependence of $n$ and $\extK$ is understood.
This follows also from the Mainardi equation \bref{mainardi2}. 
Since from  \bref{gauss} and \bref{codazzi} we have
\[
2\RicV(n,n) = {}-\RV + \RM  - \|\extK\|^2 + (\trM \extK)^2,
\]
and since $ -\gV(Y,n) =N$ is the lapse if $Y=\pd{}{s}$ is the
evolution vector of a foliation by spacelike hypersurfaces, we find
that \bref{eq:2var} can be rewritten as
\begin{equation}
   \label{eq:2varV}
   \RV = \RM + \|\extK\|^2 + (\trM \extK)^2 - 2 N^{-1}\Delta_M N +2  
D_n(\trM  \extK)\;.
\end{equation}
Integrating this quantity over the region bounded by two spacelike  
hypersurfaces $F( \{s_0 \}\times M)$ and  $F( \{s_1 \}\times M)$, we  
obtain the Lagrangian\index{Einstein-Hilbert Lagrangian}
\begin{eqnarray}\nonumber
   \cL(g,\extK)  &=& \int_{F([s_0,s_1]\times M)} \RV\,dv_V
   \\ \nonumber
   &=& \int_{s_0}^{s_1}\int_{M_s} \left(\RM + \|\extK\|^2 - (\trM
   \extK)^2\right) N - 2
   \Delta_M N \, dv_{\gM}\ ds
   \\
   \label{eq:2}
   && + \int_{M_{s_1}-M_{s_0}} 2\trM \extK\, dv_{\gM}\;,
\end{eqnarray}
using \bref{eq:19}.  From this Lagrangian, we calculate $\pi^{ij}$, 
the momentum conjugate to ${\gM}_{ij}$ (see \cite{Wald84}) as
\begin{equation}
   \label{eq:1}
   \pi^{ij} =  \frac{\delta \cL}{\delta \dot{\gM}_{ij}} =
   \sqrt{\det\gM}\,(\extK^{ij}-\trM
   \extK \gM^{ij})\;.
\end{equation}
Then ignoring boundary terms, we calculate the Hamiltonian
$\cH=\int _M\pi^{ij}\dot{\gM}_{ij}-\cL$  in ADM form: \index{ADM!Hamiltonian}
\begin{equation}
   \label{eq:3}
   \cH_{ADM} = -\int_M \xi^\al \Phi_\al(g,\pi), 
\end{equation}
where $\xi^\al=(N,X^i)$ is the four-vector consisting of the lapse and  
the shift, and where \index{constraint operators}
\begin{eqnarray}
   \label{Phi0}
   \Phi_0 &=&  \RM \det(\gM)^{1/2} - \det(\gM)^{-1/2} (||\pi||^2 -\half  
(\trM \pi)^2 )
\\
\label{Phii}
    \Phi_i &=&  {}-2 \gM_{ij}\gradM_k\pi^{jk}
\end{eqnarray}
are the constraint operators, written in densitized form, in terms of
the canonical variables $(\gM,\pi)$ rather than $(\gM,\extK)$.  
It is easily verified that the equations $\Phi_a(\gM,\extK)=0$,
$a=0,\dots,3$ are equivalent to \bref{Hcon},\bref{pcon} with $T_{0a}=0$.

Given an explicit expression for the Hamiltonian such as
\bref{eq:3}, we can readily calculate the evolution equations for
$(\gM,\pi)$ and for any functional of these quantities. In particular,
we have the ADM form \cite{ADM62} of the evolution equations 
\index{ADM!dynamical equations}
\bref{eq:dtga}, \bref{eq:dtK}  
\begin{equation}
   \label{ADM}
   \frac{d}{dt} \left[
     \begin{array}{c} \gM \\ \pi  \end{array} \right]
   =  \left[ \begin{array}{cc} 0&1\\-1&0 \end{array} \right]
    D\Phi_\al(\gM,\pi)^*\xi^\al,
\end{equation}
where $D\Phi_\al^*$ is the formal adjoint of the linearization (or
functional derivative operator) $D\Phi_\al$.  Explicitly we have
\index{constraint equations!linearized}
\cite{FischerMarsden79}:
\begin{eqnarray}
   \lefteqn{D\Phi(\gM,\pi)(h,p)\ \ = } \qquad       
\nonumber\\
 && \biggl( \left(\delta\delta h - \Delta\trM h
     - h_{ij}\left( \RicM{}^{ij}- \half \RM\gM^{ij}\right)\right)\sqrt{\gM}
\nonumber \\
 &&{}+ h_{ij}\left( \trM\pi\,\pi^{ij} - 2\pi^i_k\pi^{kj}
                  +\half |\pi|^2 \gM^{ij} 
                  -\tfrac{1}{4}(\trM\pi)^2\gM^{ij}
                  \right)/\sqrt{\gM}                                  
\nonumber \\   
 &&{}+ p^{ij}\left(\trM\pi \gM_{ij}-2\pi_{ij}\right)/\sqrt{\gM}\ ,
\nonumber \\
 &&{}       \pi^{jk}\left(2\nabla_j h_{ik} - \nabla_i h_{jk}\right)
      + 2 h_{ij}\nabla_k \pi^{jk}  + 2\gM_{ik}\nabla_j p^{jk} \biggr),
\label{DPhi}
\end{eqnarray}
\begin{eqnarray}
\lefteqn{D\Phi(\gM,\pi)^*\xi \ \ =} \qquad 
\nonumber \\
 &&    \biggl( \left(\nabla^{i}\nabla^jN - \Delta N\gM^{ij}
     - N  \RicM{}^{ij} + \half N \RM\gM^{ij}\right)\sqrt{\gM} 
 \nonumber \\
 &&{}+ N \left( \trM\pi\,\pi^{ij} - 2\pi^i_k\pi^{kj}
                    +\half |\pi|^2 \gM^{ij} 
                    -\tfrac{1}{4}(\trM\pi)^2\gM^{ij}
        \right)/\sqrt{\gM}
\nonumber \\ 
   && {}    +\left( X^k\nabla_k\pi^{ij} + \nabla_kX^k \pi^{ij}
               -2\nabla_kX^{(i}\pi^{j)k} \right)\ ,
\nonumber \\
 &&{}   N \left(\trM\pi \gM_{ij }-2\pi_{ij}\right)/\sqrt{\gM}
        - 2 p^{ij}\nabla_{(i}X_{j)}   \biggr),
\label{DPhi*}
\end{eqnarray}
where $\delta\delta h = \nabla^{i}\nabla^jh_{ij}$ and $\xi=(N,X^i)$.

Note that in the literature there are a number of different
explicit expressions for the time derivatives of $\gM$ and $\pi$
\cite{AndersonYork99}. These expressions can all be related by the
addition of terms which vanish if the constraint equations are
satisfied.

The adjoint used in \bref{ADM} is formal, in that it ignores the
contributions of boundary terms. For spatially compact spacetimes,
there are of course no boundary terms. For asymptotically flat
spacetimes, the resulting asymptotic boundary terms in the Hamiltonian
are closely related to the ADM total energy-momentum \cite{Wald84},
and they play a major role in the analysis of the physics of such
spacetimes.

An important consequence of the form \bref{ADM} of the evolution
equations is the result of Moncrief \cite{Moncrief75},
which shows that a vector field $\xi=(N,X^i)$ satisfies
$D\Phi(\gM,\extK)^*\xi=0$ for sufficiently smooth vacuum initial
data $(\gM,\extK)$ if and only if $\xi$ is the restriction to $M$ of
a Killing field in the spacetime development of the initial data.

\subsection{Preserving the Constraints}
\label{sec:prescon}

If $(\gM_0, \pi_0)$ is initial data which satisfies the (vacuum)
constraint equations and has vacuum evolution $(\gM(t), \pi(t))$
corresponding to some choice of lapse-shift $(N,X^i)$, it is natural
to ask if the evolving data satisfies the constraints for all values
of $t$.  A first step in showing this is to calculate the time
derivative of the constraint operators $H \equiv \Phi_0$ and $J_i
\equiv \Phi_i$, cf.  \cite{AndersonYork98}:
\begin{eqnarray}
   \label{Hdot}
   \pd{}{t}H &=& \nabla_m (X^m H) +2\gM^{ij} \nabla_i N  J_j   + N  
\nabla_i J^i +\tfrac{1}{2} N \gM^{-1/2} H \gM_{ij} \pi^{ij} \;,
   \\
   \label{Jdot}
   \pd{}{t}J_i &=& 2 \nabla_i N H +N \nabla_i H +\nabla_k (X^k J_i)  
+\nabla_i X^k J_k\;.
\end{eqnarray}
These formulas follow directly from the conservation law
$\nabla^bG_{ab}=0$ and the fact that the evolution equation
\bref{eq:dtK} implies $G_{ij}=0$, $1\le i,j\le 3$.  The virtue of
these expressions for the time derivatives of the constraints 
is that together they comprise a symmetric hyperbolic system. We may
then use standard energy arguments (based on an energy quantity which
is quadratic in $H$ and $J_i$) to argue that if initially $H$ and
$J_i$ vanish, then for all time consistent with a foliation of the
spacetime development of the initial data, $H$ and $J_i$ vanish as
well.  \index{constraint equations!preserved under ADM evolution}


The formal Hamiltonian interpretation of the evolution equations
\bref{ADM} regards $(\gM(t), \pi(t))$ as defining a trajectory in the
cotangent bundle $T^*\cM$ of the space $\cM$ of Riemannian metrics
$\gM_{ij}$ on $M$. It is straightforward to endow this phase space
\index{Hamiltonian phase space}
$\cF$ with the structure of a Hilbert or Banach manifold, for example
$(\gM_{ij},\pi^{ij})\in H^s(M)\times H^{s-1}(M)$, $s\ge 2$ for compact
$M$ using the Sobolev spaces $H^s$, and then the evolution \bref{ADM}
is determined from a \emph{densely defined} vector field on $\cF$.
The fact that the constraints are preserved shows that a trajectory
of the evolution \bref{ADM} lies in the constraint set $\cC$, the
subset of $\cF$ satisfying the constraint equations, provided the
initial data $(\gM(0),\pi(0))$ lies in $\cC$.

Note that while the constraints are preserved under the explicit,
smooth, evolution generated by \bref{ADM}, if we evolve instead
numerically (with the consequent unavoidable numerical errors), then
the dynamical trajectory will leave the constraint set $\cC$.  Indeed,
numerical tests suggest that such a trajectory will exponentially
diverge from $\cC$. This property, and how to control it, is under
intensive study by numerical relativists; for a recent example see
\cite{LindblomScheel03}.

\subsection{Linearization Stability}
\label{sec:linearization}
\index{linearization stability}
The linearized Einstein equations govern the evolution of perturbations
$\delta\gV$ of a given background spacetime metric $\gV_0$ and have
been extensively analysed, in view of their importance in astrophysics
and cosmology.  It is then natural to ask whether linearized solutions
actually correspond to solutions to the full nonlinear equations; that
is, whether a given solution of the linearized equations is tangent to
some curve of solutions of the full equations.  Such a perturbation
solution is called \emph{integrable}.  By the fundamental local
existence theorem \cite{FouresBruhat52} of Yvonne Choquet-Bruhat, a
perturbation is integrable if all solutions of the linearized
constraint equations $D\Phi_{(\gM,\pi)}(h,p)=0$ correspond to full
solutions of the constraints; i.e., ~if the constraint set $\cC$
is a Hilbert submanifold.  This is the question of \emph{linearization
  stability} and it is somewhat surprising that this fails in certain
situations.

The implicit function theorem (see e.g. \cite{AbrahamEtal88}) shows
that $\cC$ is a Hilbert submanifold of $\cF$ provided the
linearization $D\Phi_{(\gM,\pi)}$ has closed
range and is surjective. Now the range of $D\Phi_{(\gM,\pi)}$ is
$L^2(M)$-orthogonal to the kernel of the adjoint
$D\Phi_{(\gM,\pi)}^*$, which corresponds by work of Moncrief
\cite{Moncrief75,Moncrief76} to space-time Killing vectors.  Using
this insight and elliptic theory for the operator
$D\Phi_{(\gM,\pi)}D\Phi_{(\gM,\pi)}^*$, it has been shown by Fischer
and Marsden \cite{FischerMarsden79,FischerEt80,ArmsMarsdenMoncrief82}
that if the data $(\gM,\pi)$ does not admit Killing vectors and if $M$
is compact then $\cC$ is locally a Hilbert manifold, at least near
smoother data.  

Moreover, at points of $\cC$ corresponding to data for spacetimes with
a Killing vector field, $\cC$ is not a submanifold but instead instead
has a cone-like singularity arising from a quadratic relation
\cite{FischerEt80}.  This condition arises from the second derivative
of a curve in $\cC$, $\lambda \mapsto \Phi(\gM(\lambda),\pi(\lambda)$,
about $(\gM_0,\pi_0)$ admitting a Killing field $\xi$.  Integrating
over $M$, discarding boundary terms and using
$D\Phi_{(\gM,\pi)}^*\xi=0$ gives
\begin{equation}
\label{quad}
  \int_M \xi\cdot D^2\Phi_{(\gM_0,\pi_0)}((h,p),(h,p))\,d^3x = 0,
\end{equation}
which is the required quadratic condition on the constrained
variations $(h,p)$.  In \cite{Moncrief76} it is shown that this
condition is equivalent to the vanishing of the Taub quantity,
constructed from a solution of the linearized Einstein equations about
a spacetime admitting a Killing vector.

The situation for asymptotically flat spacetimes differs significantly
from the case of compact $M$.  Using weighted Sobolev spaces (with
local regularity $\gM\in H^s$, $s=2$) it is possible to show that the
$\cC$ is everywhere a Hilbert submanifold \cite{Bartnik04a}; although
Killing vectors such as rotations are possible, they do not satisfy
the asymptotic conditions needed to prevent surjectivity of
$D\Phi_{(\gM,\pi)}$ and thus the quadratic conditions on constrained
variations do not arise.  This work \cite{Bartnik04a} also shows that
the asymptotically flat phase space $\cF = (\go+ H^2_{-1/2})\times
H^1_{-3/2}$ \index{Hamiltonian phase space}
provides a natural setting in which the ADM Hamiltonian and
energy-momentum functionals become smooth functions, for example.


\section{Solving the Constraint Equations}

There are three reasons for seeking to construct and study solutions
of the Einstein constraint equations. First, we would like to obtain
initial data sets which model the initial states of physical systems.
Evolving such data, we can model the gravitational physics of those
physical systems. Second, we would like to understand the space of all
globally hyperbolic solutions of Einstein's gravitational field
equations. Since, as noted earlier, Einstein's equations are
well-posed, the space of solutions of the constraints parametrizes the
space of solutions of the spacetime field equations. Third, we would
like to know enough about the nature of the space of solutions of the
constraints and the Hamiltonian dynamics of the classical Einstein
equations on this space to be able to consider descriptions of
gravitational physics which are consistent with the quantum principle
(i.e., to ``quantize gravity'').

The easiest way to find a set of initial data $(M^3, \gM, \extK)$ which
satisfies the Einstein constraint equations is to first find an
explicit solution $(V^4, \gV)$ of the spacetime Einstein equations
(e.g., the Schwarzschild or Kerr solutions), and then to choose a
Cauchy surface $M^3 \hookrightarrow V^4$ in that spacetime solution:
The induced Riemannian metric $\gM$ and second fundamental form  $\extK$
on $M^3$ together solve the constraints. Since the set of known
spacetime solutions is very limited, this procedure is not especially
useful for making progress towards the goals discussed above (although
we shall see below that solutions of the constraints obtained from
slices  of known spacetime solutions can be very handy as building
blocks for ``gluing constructions''  of solutions of the constraints,
which are in turn potentially very useful for modeling physical
systems). 

Our focus here is on more comprehensive methods for finding and
studying solutions of the constraints. We first discuss the conformal
method. This is by far the most widely used approach, both for
physical modeling and for mathematical analyses, and so it receives
the bulk of our attention here. We next consider the thin sandwich and
conformal thin sandwich ideas. The latter is closely related to the
conformal method; we compare and contrast the two here. The fourth
approach we consider is the quasi-spherical ansatz. Although not as
comprehensive as the conformal method, the quasi-spherical ansatz is
potentially useful for certain key applications, which we discuss
below. We conclude by describing the recently developed procedures for
gluing together known solutions of the constraints, thereby producing
new ones.

For convenience, we work primarily in this paper with the vacuum
constraint equations. We note, however, that most of what we discuss
here generalizes to the Einstein-Maxwell, Einstein-Yang-Mills,
Einstein-fluid, and other such nonvacuum field theories.


\subsection{The Conformal Method}

The vacuum Einstein constraint equations \bref{Hcon}, \bref{pcon} with
$T_{nn}=0$ and $T_{ni}=0$ constitute an underdetermined system of four
equations to be solved for twelve unknowns $\gM$ and $\extK$. The idea
of the conformal method is to divide the initial data on $M^3$ into
two sets---the ``Free (Conformal) Data'',\index{free conformal data}
and the ``Determined Data''---in such a way that, given a choice of
the free data, the constraint equations become a {\em determined}
elliptic PDE system to be solved for the Determined Data.

There are at least two ways to do this (see \cite{Cook03} for others).
The sets of Free Data and Determined Data are the same for both
procedures; we have
\begin{description}
\item[Free (``Conformal'') Data:]\ %
\par

\begin{tabular}{rcl}
  $\lambda_{ij}$ &---& a Riemannian metric,
specified up to conformal factor;
\\
$\sigma_{ij}$ &---& a divergence-free\footnotemark
($\nabla ^i\sigma_{ij} = 0$),
trace-free ($\lambda^{ij}\sigma_{ij} =0$)
\\
&& symmetric tensor;
\\ $\tau$ &---& a scalar field,
\end{tabular}
\footnotetext{In the free data,
the divergence-free condition is defined using the Levi-Civita covariant
derivative compatible with the conformal metric $\lambda_{ij}$.}%
\item[Determined Data:]\ %
\par

\begin{tabular}{rcl}
$\phi$ &---& a positive definite scalar field,
\\
   $W^i$ &---& a vector field.
\end{tabular}
\end{description}
The difference between the two procedures has to do with  the form of
the equations to be solved for the Determined Data, and the way in
which the two sets of data are combined to obtain  $\gM$ and $\extK$
satisfying the constraints. In one of the procedures, which we label
the ``Semi-Decoupling Split'' (historically called ``Method
A''), the equations for $W$ and $\phi$ take the form \\

\index{Lichnerowicz equation}
\noindent
{\textbf{Semi-Decoupling Split}}
\begin{eqnarray}
\nabla_i (LW)^i_j &=& \tfrac{2}{3}\phi^6\nabla_j\tau
\label{confmomA}
\\
\Delta\phi &=& \tfrac{1}{ 8}R\phi - \tfrac{1}{ 8}(\sigma^{ij} +
LW^{ij})(\sigma_{ij} + LW_{ij})\phi^{-7}  + \tfrac{1}{ 12}\tau^2 \phi^5,
\label{LichneroA}
\end{eqnarray}
where the  Laplacian $\Delta$ and the scalar curvature $R$ are
based on the $\lambda_{ij}$-compatible covariant derivative
$\nabla_i$, where $L$ is the corresponding conformal Killing
operator, defined by
\begin{equation}
\label{LW}
(LW)_{ij}\equiv \nabla_iW_j +\nabla_jW_i -\tfrac{2}{3} \lambda_{ij} 
\nabla_k W^k,
\end{equation}
  and we construct $\gM$ and
$\extK$ from the free and the determined data as follows
\begin{eqnarray}
\gM_{ij} &=& \phi^4\lambda_{ij}
\label{recongammaA}
\\
\extK_{ij} &=& \phi^{-2}(\sigma_{ij }+LW_{ij}) + \tfrac{1}{
   3}\phi^{4}\lambda_{ij}\tau.
\label{reconKA}
\end{eqnarray}
Using the other procedure, which we label the ``Conformally Covariant
Split'' (historically ``Method B''), the equations for $W$ and $\phi$
are\\

\noindent{\textbf{Conformally Covariant Split}}
\begin{eqnarray}
\nabla_i (LW)^i_j &=& \tfrac{2}{3}\nabla_j\tau -
   6(LW)^i_j\nabla_i \ln \phi
\label{confmomB}
\\
\Delta\phi&=&\tfrac{1}{ 8}R\phi-\tfrac{1}{
   8}\sigma^{ij}\sigma_{ij}\phi^{-7}-\tfrac{1}{
   4}\sigma^{ij}(LW)_{ij}\phi^{-1} \nonumber
\\
   &&{}+\tfrac{1}{12}(\tau^2-(LW)^{ij}(LW)_{ij})\phi^5
\label{LichneroB}
\end{eqnarray}
and the formulas for $\gM$ and $\extK$ are
\begin{eqnarray}
\gM_{ij} &=& \phi^4\lambda_{ij}
\label{recongammaB}
\\
\extK_{ij} &=& \phi^{-2}\sigma_{ij }+\phi^4 LW_{ij} + \tfrac{1}{
   3}\phi^{4}\lambda_{ij}\tau.
\label{reconKB}
\end{eqnarray}

Each of these two methods has certain advantages. For the
semi-decoupling split, if one chooses the mean curvature $\tau$ to be
constant, then the $\phi$ dependence drops from \bref{confmomA}, and
the the focus of the analysis is on \bref{LichneroA}, the Lichnerowicz
equation. This is not true for the conformally covariant split;
however in this latter case, one finds that a solution exists for free
data $(\lambda_{ij}, \sigma_{ij}, \tau)$ if and only if a solution
exists for $(\theta^4 \lambda_{ij}, \theta^{-2}\sigma_{ij}, \tau)$.
Far more is known mathematically about the semi-decoupling split, and
this approach has been used much more in applications, so we focus on
the semi-decoupling split in the rest of this paper. We do note,
however, that numerical relativists have very recently begun to apply
the conformally covariant split for certain studies \cite{Cook03}.

  Is it true that, for every choice of the free data $(\lambda_{ij},
  \sigma_{ij}, \tau)$, one can always solve equations  \bref{confmomA}
  and \bref{LichneroA} for the determined data $(\phi, W)$ and thereby
  obtain a solution of the constraint equations? It is easy to see that
  this is not the case: Let us choose the manifold $M^3$ to be the
  three sphere, and on $M^3$ we choose a metric $\lambda$ with
  non-negative scalar curvature, we choose $\sigma$  to be zero
  everywhere, and we choose $\tau$ to be unity everywhere. One readily
  verifies that every solution to the equation $\nabla_i (LW)^i_j =0$
  has $LW_{ij}=0$. The Lichnerowicz equation then becomes $\Delta\phi=R\phi
  +\phi^5$. Since the right hand side of this equation is positive
  definite (recall the requirement that $\phi>0$), it follows from
  the maximum principle on closed (compact without boundary) manifolds
  that there is no solution.

  In light of this example, the main question is really the following:
  For which choices of the manifold $M^3$ and the free data can one
  solve \bref{confmomA} and \bref{LichneroA}? It turns out that we know
  quite a bit about the answer to this question, yet still have quite a
  bit to learn as well. To describe what we know and do not know, it is
  useful to categorize the question using the following criteria:

  \begin{description}
\item[Manifold and Asymptotic Conditions]\
\begin{itemize}
\item Closed
\item Asymptotically Euclidean
\item Asymptotically hyperbolic
\item Compact  with boundary conditions
\item Asymptotically Euclidean with interior boundary conditions
\end{itemize}
\item[Regularity]\
\begin{itemize}
\item Analytic
\item Smooth
\item Sobolev and Holder classes
\item Weak
\item Asymptotic fall off conditions at infinity
\end{itemize}
\item[Metric Conformal Classes]\
\begin{itemize}
\item Yamabe positive
\item Yamabe zero
\item Yamabe negative
\end{itemize}
\item[Mean Curvature]\
\begin{itemize}
\item Constant (``CMC'')
\item Near Constant (small $\frac{\max |\nabla \tau|}{\min |\tau| }$)
\item Non Constant
\end{itemize}
\item [$ \sigma$ Classes]\
\begin{itemize}
\item $\sigma$ is zero everywhere on $M^3$
\item $\sigma$ not zero everywhere on $M^3$
\end{itemize}
\end{description}

The mean curvature of the data turns out to be the most important
factor in separating those sets of free data for which we know whether
or not  a solution exists from those sets for which we do not. In fact,
if the mean curvature is either constant or near constant, we can
almost completely determine whether or not a solution exists (at least in 
those cases with no boundaries present). On the other hand, if
the mean curvature is neither constant nor near constant, we know very
little. For sufficiently smooth free data, we may summarize the known
results as follows:

\index{constant mean curvature!conformal construction}
\begin{description}
\item[Constant Mean Curvature]\
\begin{itemize}
\item {\it Closed}: Completely determined. For all but certain special
   cases (see example discussed above), solutions exist. See
   \cite{ChoquetYork79} and \cite{Isenberg95}.
\item {\it {Asymptotically Euclidean}}: Completely determined. For all
   but certain special cases (depending entirely on the conformal
   class), solutions exist. See \cite{Cantor77} and \cite{BrillCantor81}.
\item {\it {Asymptotically Hyperbolic}}: Completely
   determined. Solutions always exist. See
   \cite{AnderssonChruscielFriedrich92} and
   \cite{AnderssonChrusciel96}
\item {\it{Compact with Boundary}}: Some cases determined, most  
unresolved.
\item {\it {Asymptotically Euclidean with Interior Boundary}}: Some 
cases determined, some unresolved. See
    \cite{Maxwell03} and
    \cite{Dain03}
\end{itemize}
\item[Near Constant Mean Curvature]\
\begin{itemize}
\item {\it Closed}: Mostly determined. For all but certain cases,
  solutions exist. One special case known of nonexistence. A small
  number of special cases unresolved. See \cite{IsenbergMoncrief96},
  \cite{Isenberg03} and
  \cite{IsenbergOMurchadha03}
\item {\it {Asymptotically Euclidean}}: Mostly determined. No cases
   known of nonexistence. Some cases unresolved. See
   \cite{ChoquetIsenbergYork00}
\item {\it {Asymptotically Hyperbolic}}: Mostly determined. No cases
   known of nonexistence. A small number of cases unresolved. See
   \cite{IsenbergPark97}
\item {\it{Compact with Boundary}}: Small number of cases determined,
   most unresolved.
\item {\it {Asymptotically Euclidean with Interior Boundary}}: Nothing 
determined.
\end{itemize}
\item[Non Constant Mean Curvature]
Very little known.
\end{description}

This summary is the compilation of  results from a  large number of
works (note references listed above), including some not yet written
up. The summary is fairly sketchy, since there is inadequate space
here to include many of the details. We do, however, wish to give a
flavor of what these results say more precisely, and how they are
proven. To do this we shall discuss a few representative sub-cases.

We first consider the case in which the manifold is presumed closed,
and in which we choose free data with constant $\tau$. As noted above,
as a consequence of the CMC condition, the equation \bref{confmomA}
for $W$ is readily solved, and we have $LW=0$. What remains is the
Lichnerowicz equation, which takes the form
\begin{equation}
\Delta \phi = \tfrac{1}{ 8}R\phi - \tfrac{1}{ 8}\sigma^{ij}
\sigma_{ij}\phi^{-7} + \tfrac{1}{ 12}\tau^2 \phi^5.
\label{Lichnero}
\end{equation}

There are three key PDE analysis theorems which allow us to determine,
for any set of free data consistent with these conditions, whether or
not a solution to the Lichnerowicz equation exists. We state each of
these theorems in a form most suited for this purpose (See
\cite{Isenberg95} for much more detail):
\begin{enumerate}
\item {\it The Maximum Principle:}
If we choose free data such that for any  $\phi >0$ the right hand
side of the Lichnerowicz equation \bref{Lichnero} is either positive
definite or negative definite, then (with $M^3$ presumed closed) for
that free data there can be no solution.
\item
  {\it The Yamabe Theorem:}\index{Yamabe theorem}
For  any given Riemannian metric $\lambda$ on a closed three manifold,
there is a conformally related metric $\theta^4 \lambda$ which has
constant scalar curvature. For fixed $\lambda$, all such
conformally-related constant scalar curvature metrics have the same
sign for the scalar curvature. This allows one to partition the set of
all Riemannian metrics on $M^3$ into three classes, depending on this
sign. We label these {\em Yamabe classes} as follows: ${\mathcal
   Y}^+(M^3)$, ${\mathcal Y}^0(M^3)$, and ${\mathcal Y}^-(M^3)$.
\item \index{sub- and super-solutions}
{\it The Sub and Super Solution Theorem:} If there exist a pair of
positive functions $\phi_{+}>\phi_{-}$ such that
\begin{equation}
\Delta\phi_{+} \leq \tfrac{1}{ 8}R\phi_{+} - \tfrac{1}{ 8}\sigma^{ij}
\sigma_{ij}\phi_{+}^{-7}  + \tfrac{1}{ 12}\tau^2 \phi_{+}^5
\label{super}
\end{equation}
and
\begin{equation}
\Delta\phi_{-} \geq \tfrac{1}{ 8}R\phi_{-} - \tfrac{1}{ 8}\sigma^{ij}
\sigma_{ij}\phi_{-}^{-7} + \tfrac{1}{ 12}\tau^2 \phi_{-}^5,
\label{sub}
\end{equation}
then there exists a solution $\phi$ of the Lichnerowicz equation
\bref{Lichnero}, with $\phi_{+}\geq \phi \geq \phi_{-}.$
\end{enumerate}

The Yamabe Theorem is very useful because the Lichnerowicz equation in
the form \bref{Lichnero} (with $LW=0$) is conformally covariant in the
sense that it has a solution for a set of free data $(\lambda, \sigma,
\tau)$ if and only if some function $\psi>0$ it has a solution for the
free data  $(\psi^4 \lambda, \psi^{-2}\sigma,  \tau)$. Hence, to check
solubility for a given set of free data, we can always first perform a
preliminary conformal transformation on the data and therefore work
with data for which the scalar curvature is constant.

It is now straightforward to use the Maximum Principle (together with
the Yamabe Theorem) to show that  there is a collection of classes of
CMC free data for which no solution exists. This holds for all of the
following classes of free data:
\begin{eqnarray*}
(\lambda \in {\mathcal Y}^+(M^3),
\sigma \equiv 0, \tau =0),  &\qquad&
(\lambda \in {\mathcal Y}^+(M^3), \sigma \equiv 0, \tau \neq 0),\\
  (\lambda \in {\mathcal Y}^0 (M^3), \sigma \equiv 0, \tau \neq 0),   
&\qquad&
  (\lambda \in {\mathcal Y}^0 (M^3), \sigma \neq 0, \tau =0), \\
  (\lambda \in {\mathcal Y}^- (M^3), \sigma \equiv 0, \tau =0),    
&\qquad&
  (\lambda \in {\mathcal Y}^- (M^3), \sigma \neq 0, \tau \neq 0).
\end{eqnarray*}
  Note that here,``$\sigma \neq 0$''  means that the tensor $\sigma$ is
  not identically zero on $M^3$.

  It is true, though not so simple to show, that for all other classes
  of CMC free data on a closed manifold, a solution does exist. We show
  this using the sub and super solution theorem.  For free data of the
  type $(\lambda \in {\mathcal Y}^- (M^3), \sigma \equiv 0, \tau \neq
  0)$, it is relatively easy to show that there are constant sub and
  super solutions. We merely need to find a constant $\phi_+$
  sufficiently large so that $-\phi_{+} + \tfrac{1}{ 12}\tau^2
  \phi_{+}^5$ is everywhere positive, and a constant $\phi_-$
  sufficiently small (yet positive) so that $-\phi_{-} +
  \tfrac{1}{12}\tau^2 \phi_{-}^5$ is everywhere negative. Constant sub
  and super solutions are also readily found for free data of the type
  $(\lambda \in {\mathcal Y}^- (M^3), \sigma \neq 0, \tau \neq 0)$, or
  $(\lambda \in {\mathcal Y}^0 (M^3), \sigma \equiv 0, \tau = 0)$.  For
  the remaining types of free data, we need to find either a
  nonconstant sub solution or a nonconstant super solution. This takes
  a bit of work; we show how to do it in \cite{Isenberg95}.

It is worth noting that the proof of the sub and super solution
theorem (See \cite{Isenberg95}) is a constructive one: Starting with the
super solution $\phi_+$, one solves a sequence of linear equations,
and one then shows that the monotonic sequence of solutions $\phi_+
\geq \phi_1 \geq \phi_2 \geq \dots \geq \phi_-$ converges to a solution
of the Lichnerowicz equation.  This constructibility can be useful for
numerical relativity, as has been shown in \cite{Garfinkle03}.

We next consider the case in which the free data is chosen to be
asymptotically Euclidean (See, for example,
\cite{ChoquetIsenbergYork00} for a definition of this property), with
$\tau \equiv 0$. For asymptotically Euclidean metrics, a Yamabe type
theorem again plays a key role. Proven by Brill and Cantor
\cite{BrillCantor81}, with a correction by Maxwell \cite{Maxwell03}, 
this result says that if $\lambda$ is
asymptotically Euclidean, and if for every nonvanishing, compactly
supported, smooth function $f$ we have
\begin{equation}
\inf_{ \{ f \not\equiv 0\} }  \frac{\int_M (|\nabla f|^2 + R f^2) \sqrt{\det 
\lambda}}{||f||^2_{L^{2*}}}>0,
\label{YamabeAE}
\end{equation}
then for some conformal factor $\theta$, the scalar curvature of
$\theta^4 \lambda$ is zero. Moreover, if an asymptotically Euclidean
metric fails to satisfy this condition, then there is no such
transformation. Metrics which do satisfy this condition have been
 labeled (somewhat misleadingly) as ``positive Yamabe metrics''.

The positive Yamabe property just defined is exactly the condition
which determines if, for a set of asymptotically Euclidean free data,
the Lichnerowicz equation admits a solution. That is, as proven in
\cite{Cantor77}, \bref{Lichnero} admits a solution $\phi$ with suitable
asymptotic properties if and only if the (asymptotically Euclidean)
free data $(\lambda, \sigma, \tau=0)$ has positive Yamabe metric
$\lambda$. This is true regardless of whether $\sigma$ vanishes or
not. This result is not proven directly using a sub and super solution
theorem, but the proof does involve a converging sequence of solutions
of linear equations, and is therefore again constructive.

The last case we consider here is that of near constant mean curvature
free data on closed manifolds. This case is more complicated than the
CMC case, since now we must work with the coupled system
\bref{confmomA}-\bref{LichneroA}. However, with sufficient control
over the gradient of $\tau$, we can handle the coupled system, and
determine whether or not solutions exist for almost all sets of near
CMC free data.

The key extra tool we use to carry out this analysis is the elliptic
estimate for the ``vector Laplacian'' operator $\nabla_iL(\
)^i_j$ appearing on the left side of \bref{confmomA}. Such a result
holds for any linear, elliptic, invertible operator \cite{Besse87} such
as this one.\footnote{The operator $\nabla_iL(\ )^i_j$ is invertible only if
$\lambda$ has no conformal Killing field. However, the pointwise
estimate \bref{LWestimate} we obtain for $|LW|$ can be derived even if
$\lambda$ has a conformal Killing field, using a slightly more
complicated elliptic estimate than the one presented here.}
One has the {\it Elliptic Estimate:} $||W^j||_{H_{k+2}} \leq
C ||\nabla_i(LW)^i_j||_{H_k}$, where $H_k$ is the Sobolev space of
square integrable vector fields for which the first $k$ derivatives
are square integrable as well, where $||\;||_{H_k}$ is the
corresponding norm, and where $C$ is a constant depending on the
chosen geometry $(M^3, \lambda_{ij})$.  Based on this estimate,
together with the Sobolev embedding theorem \cite{Besse87} and standard
integration inequalities, we obtain a {\em pointwise} estimate of the
form
\begin{equation}
|LW| \leq \tilde{C} \max_{M^3}\phi^6 \max_{M^3} |\nabla \tau|,
\label{LWestimate}
\end{equation}
where $ \tilde{C}$ is also a constant depending only on the geometry
$(M^3, \lambda_{ij})$. This pointwise inequality \bref{LWestimate} is
the crucial estimate which allows us to handle the coupled system, for
sufficiently small $|\nabla \tau|$.

To see how to prove that a solution exists in one of these near CMC
cases, let us consider free data with $\lambda \in {\mathcal
   Y}^-(M^3), \tau>0$, and with $\frac{\max_{M^3} |\nabla
   \tau|}{\min_{M^3} |\tau| }$ sufficiently small.\footnote{The
   statement that ``$\frac{\max_{M^3} |\nabla \tau|}{\min_{M^3} |\tau|
   }$ (is) sufficiently small'' may seem nonsensical, since this
   quantity is dimensional. However, as seen in the proof
   \cite{IsenbergMoncrief96}, the more precise statement of this
   condition involves the constant $C$, which has dimensions as well.}
We claim that, for such free data, a solution exists. To show this, we
consider the sequence of semi-decoupled PDE systems
\begin{eqnarray}
\nabla_i (LW_{(n)})^i_j &=& \tfrac{2}{3}\phi_{(n-1)}^6\nabla_j\tau
\label{confmomA-n}
\\
\Delta\phi_{(n)}& =& \tfrac{1}{ 8}R\phi_{(n)} - \tfrac{1}{ 
8}(\sigma^{ij} +
LW_{(n)}^{ij})(\sigma_{ij} + LW_{(n)ij})\phi_{(n)}^{-7}  + \tfrac{1}{
   12}\tau^2 \phi_{(n)}^5.
\label{LichneroA-n}
\end{eqnarray}
Choosing a suitable initializing value for $\phi_0$, we first show
that a sequence $(\phi_{(n)}, W_{(n)})$ of solutions to
(\ref{confmomA-n}-\ref{LichneroA-n}) exists. For the decoupled
equation \bref{confmomA-n}, the existence of a solution $W_{(n)}$
follows from the invertibility of the vector Laplacian. For
\bref{LichneroA-n}, existence follows from the sub and super solution
theorem, since we readily find a sequence of constant sub and super
solutions for this choice of free data. We next show that there are
(positive) uniform upper and lower bounds for the sequence
$(\phi_{(n)})$. This is where we first need to use the estimate
\bref{LWestimate} for $|LW_{(n)}|$:  Specifically, after arguing that
there is a uniform (constant) upper bound for the  $\phi_{(n)}$'s if we
can find a positive  constant $\zeta$ such that
\begin{equation}
\zeta^3\geq \frac{3}{2 \min_{M^3}\tau^2} \zeta^2+{\frac{3} {
     \min_{M^3}\tau^2}} \sigma^{ij} \sigma_{ij} +
{\frac{3} { \min_{M^3}\tau^2}} LW_{(n)}^{ij} LW_{(n)ij},
\end{equation}
we use the $|LW_{(n)}|$ estimates to show that it is sufficient to
find a positive constant $\zeta$ satisfying
\begin{equation}
\zeta^3\geq {\frac{3} {2 \min_{M^3}\tau^2}} \zeta^2+{\frac{3} {
     \min_{M^3}\tau^2}} \sigma^{ij} \sigma_{ij} +
C{\frac{\max_{M^3}|\nabla \tau|^2}{2 \min_{M^3}\tau^2}} \zeta^3
\end{equation}
We immediately see that if, for fixed $\lambda$ and $\sigma$, we
choose $C{\max_{M^3}|\nabla \tau|^2}/({2 \min_{M^3}\tau^2})$
sufficiently small, then such a $\zeta$ exists. The existence of the
uniform upper bound for $\phi_{(n)}$ follows.

Establishing these uniform bounds is the crucial step in our proof
that solutions exist for free data of the type we are discussing here,
as well as for other classes of near CMC free data. Once we have these
bounds, we can carry through a contraction mapping argument to show
that the sequence $(\phi_{(n)}, W_{(n)})$ converges (a bit more
squeezing of $|\nabla \tau|$ is needed to carry this out). We can then
go on to use continuity arguments to show that the limit is a weak
solution, and finally use bootstrap arguments to show that in fact the
weak solution is a strong solution, of the desired smoothness. The
details of this proof are presented in \cite{IsenbergMoncrief96}.

A very similar argument can be used to prove that equations
\bref{confmomA}-\bref{LichneroA} admit solutions for a number
of classes of near CMC free data, including the following:
\begin{enumerate}
\item
$\lambda \in {\mathcal Y}^-(M^3), \tau < 0$, and ${\frac{\max_{M^3} 
|\nabla
     \tau|} {\min_{M^3} |\tau|}}$ sufficiently small;
\item
$\lambda \in {\mathcal Y}^+(M^3)$, $\sigma$ not identically zero, and
${\frac{\max_{M^3} |\nabla \tau|} {\max_{M^3} |\sigma| }}$ sufficiently
small;
\item
$\lambda \in {\mathcal Y}^0(M^3)$, $\tau$ nowhere zero, $\sigma$ not
identically zero, and ${\frac{\max_{M^3} |\nabla \tau|}{\max_{M^3}
     |\sigma| + \max_{M^3}|\nabla \tau|}}$ sufficiently small.
\end{enumerate}

What about the other near CMC cases on closed $M^3$? Just recently,
with Niall O'Murchadha, we have found for the first time a class of
near CMC free data for which a solution {\em does not} exist. The data
is of the following type:
\[
\lambda \in {\mathcal Y}^+(M^3) \cup {\mathcal Y}^0(M^3), \sigma
\equiv 0, \tau \textrm{\ nowhere\ $0$, and\ }{\frac{\max_{M^3} |\nabla 
\tau|}
     {\min_{M^3} |\tau|}} \textrm{ sufficiently small.}
\]
The proof that no solution exists for free data of this type is a
relatively simple application of the pointwise estimate
\bref{LWestimate} for $|LW|$ discussed above, and the maximum
principle. The details are presented in \cite{IsenbergOMurchadha03}.
As for near CMC data on a closed manifold which is not one of the
types discussed here, nothing is yet known. Note that, in a rough
sense, these remaining unresolved cases are not generic.

As noted in the summary above, there are a number of other classes of
free data for which we know whether or not solutions to
\bref{confmomA}-\bref{LichneroA} exist. Methods similar to those
discussed here resolve the existence question for most CMC or near CMC
free data which are asymptotically Euclidean or asymptotically
hyperbolic, as well for such data on closed manifolds. The situation
for free data specified on compact manifolds with boundary is much
less understood; however this is most likely a result of neglect
rather than difficulty. Since numerical relativity has motivated a
recent interest in this question, and since the methods discussed here
are believed to work for data on manifolds with boundary, it is likely
that in the next few years, the solvability of
\bref{confmomA}-\bref{LichneroA} for such data is likely to be
relatively well understood. Indeed, this motivation has led to the
very recent results of Maxwell \cite{Maxwell03} and Dain
\cite{Dain03}, who give sufficient conditions on asymptotically
Euclidean free data with interior ``apparent horizon'' boundary
conditions for solutions to exist

The situation is very different for non-constant mean curvature data
with no controls on $|\nabla \tau|$. Almost nothing is known, and
there are no promising techniques known at this point. New ideas are
needed.

It is worth noting that the underlying approach of specifying the mean
extrinsic curvature $\tau$ is geometrically natural, since the
converse problem, of finding a hypersurface of prescribed mean
curvature in a given spacetime, leads to a quasi-linear elliptic
equation bearing some similarities with the Euclidean minimal surface
equation.  It is known that solutions of the Lorentz mean curvature
equation are strictly spacelike and smooth, at least to the extent
permitted by the regularity of the ambient spacetime and boundary
conditions, so such prescribed mean curvature hypersurfaces provide
natural spacelike slices of the spacetime \index{constant mean
  curvature!existence}
\cite{ChoquetBruhat76,BartnikSimon82,Gerhardt83,Bartnik84,Bartnik88a}.
Although there are examples of spacetimes not admitting maximal
\index{constant mean curvature!non-existence} ($\tau=0$,
\cite{Brill82}) or CMC ($\tau=const.$, \cite{Bartnik88b},
\cite{ChruscielIsenbergPollack04}) hypersurfaces, such non-existence
behaviour is driven by global causal topology, which allows
area-maximising sequences of hypersurfaces to become unbounded to the
past or future; see \cite{Bartnik88a,Bartnik88b}.  It is therefore not
surprising that \emph{a priori} control of the interior causal
geometry is an essential ingredient in the proof of existence of
entire maximal hypersurfaces in asymptotically flat spacetimes
\cite{Bartnik84}.

To wind up this discussion of the implementation of the conformal
method to find solutions of the constraint equations, it is important
to note two facts: 1) The map from sets of free data to solutions of
the constraints is surjective. This follows from the observation that
if $(\gM, \extK)$ is a solution of the constraints, then $\phi =1,
W=0$ is clearly a solution of (\ref{confmomA}-\ref{LichneroA}) for the
free data $\lambda = \gM, \sigma = \extK - \tfrac{1}{ 3} \lambda \trK,
\tau = \trK$. 2) For all those sets of free data for which solutions
of (\ref{confmomA}-\ref{LichneroA}) are known to exist (excepting for
the very special case of $\lambda$ flat, $\sigma=0$, and $\tau=$ on
$M^3=T^3$) the solution is unique.

As a consequence of these facts, those sets of free data for which
solutions exist together parametrize the space of
constraint-satisfying initial data for Einstein's equations. It
follows that, once we determine exactly which free data sets map to
solutions, we will have made significant progress toward understanding
the ``degrees of freedom'' of Einstein's theory. 
\index{gravitational degrees of freedom}

Also as a consequence of these facts, if we seek initial data
invariant under an isometry group, it is sufficient to choose free
data with this invariance.\footnote{Note that the presence of
  symmetries generally introduces integrability conditions which must
  be satisfied by the initial data, and consequently by the free data.
  These conditions are relatively easy to handle.}

In view of the dominant role the conformal method has assumed as a
tool for mathematical as well as numerical analyses of the
constraints, it is important to point out its limitations. As noted
above, the conformal method replaces the original underdetermined PDE
system of the constraint equations (four equations to be solved for
twelve unknowns), by a determined system of elliptic character. This
is very useful for a wide variety of studies (such as the
parametrization question). However for a number of other problems, it
is a bad idea.  Consider for example a given solution of the
constraint equations on a finite radius ball. Can one smoothly extend
the solution onto a neighborhood properly containing the ball? Can it
be smoothly extended to a complete, asymptotically Euclidean
solution? Might there be a smooth extension which is identically
Schwarzschild or Kerr outside some (larger) ball? To study questions
like these, casting the constraint equations into the form of a
determined elliptic system is not at all useful. It is better to
bypass the conformal method and work with the original underdetermined
system.

\subsection{The Thin Sandwich Construction}
\index{thin sandwich!conjecture}
Some time ago, Wheeler \cite{MTW73,BaierleinEt62} discussed a possible
alternative to the initial value formulation for determining a
spacetime solution of Einstein's equations. Rather than specifying
initial data in the form of a Riemannian metric $\gM$ and a symmetric
tensor $\extK$ which satisfy the Einstein constraint equations, he
asked whether one might specify a pair of Riemannian metrics $h_1$ and
$h_2$ and seek to find a (unique?) spacetime $(V^4, g)$ in which
$h_1,h_2$ arise as the induced metrics on disjoint Cauchy surfaces in
$( V^4, g)$, and which satisfies the Einstein equations i.e.~$(V^4,
g)$ is the spacetime solution ``connecting'' $h_1$ and $h_2$. This is
the \emph{sandwich conjecture}.

Considering the analogous question in electrodynamics, however, we
find that the uniqueness part of the Maxwell version of the sandwich
conjecture is false.  Hence there is reason to believe that the
uniqueness assertion in the sandwich conjecture for the Einstein
equations may also be false.  This led Wheeler to instead propose the
\emph{thin sandwich conjecture}, which postulates that given a freely
chosen Riemannian metric $\gM$ and symmetric tensor $J$ representing
the time derivative of the metric evolution in a spacetime, then we
can use the constraints to determine a lapse function and shift vector
field so that indeed, with respect to a choice of foliation and
coordinates compatible with that lapse and shift, $J$ {\em is} the
time derivative of the metric. More explicitly, having chosen $\gM$
and $J$, we seek a lapse $N$ and shift $X$ so that if we set
\index{thin sandwich!equations}
\begin{equation}
\label{eq:TSK}
\extK_{ij}=-N^{-1}(\tfrac{1}{2}J_{ij} -X_{(i;j)}),
\end{equation}
then $(\gM,\extK)$ satisfy the constraint equations
(\ref{Hcon},\ref{pcon}) with source terms $\rho=16\pi T_{nn}$,
$S_i=8\pi T_{ni}$.  Regarding $(\gM,J,\rho,S)$ as \emph{given}
fields, the constraints give four equations for $(N,X)$, the so-called
\emph{thin sandwich} equations.

It is clear that if $(\gM,\extK)$ satisfy the constraint equations,
then $J$ is the time derivative $\partial_t\gM$ of the spatial
metric in the corresponding spacetime evolution of the data
$(\gM,\extK)$ with respect to coordinates with lapse-shift $(N,X)$.

The Hamiltonian constraint is readily solved for the lapse, giving
\index{thin sandwich}
\begin{equation}
\label{eq:lapse}
N =\sqrt{\frac{(\Gamma^i_i)^2-\Gamma^{ij}\Gamma_{ij}}{\rho-\RM}},
\end{equation}
where we have set $\Gamma_{ij}=\tfrac{1}{2}J_{ij}-X_{(i;j)}$  for
brevity.  Substituting this expression for $N$ into the momentum
constraint (\ref{pcon}) we obtain the \emph{reduced thin sandwich equations}
(``RTSE'') \index{thin sandwich!reduced equations}
\begin{equation}
   \label{eq:RTSE}
\nabla_j\left[\sqrt{\frac{\rho-\RM}{(\Gamma^k_k)^2- 
\Gamma^{kl}\Gamma_{kl}}}
     (\Gamma^{ij}-\Gamma^k_k \gM^{ij})\right]=S^i,
\end{equation}
which we view as a system of partial differential equations for
$X^i$, the components of the shift.

The linearization of \bref{eq:RTSE} is shown to be elliptic in
\cite{BartnikFodor93}, provided that the conjugate momentum $\pi^{ij}$
(recall the formula \bref{eq:1} for $\pi$) is either positive or
negative definite. This rather surprising condition is guaranteed, so
long as the condition
\begin{equation}
\label{eq:rtsecond}
\rho -\RM >0
\end{equation}
is satisfied everywhere in ${M}$ and provided that $J$ is chosen in
such a way that the formula \bref{eq:lapse} for the lapse function
satisfies
\begin{equation}
\label{eq:lapsecond}
N >0.
\end{equation}
An additional necessary condition for solvability of the RTSE is that
the equation
\begin{equation}
\label{eq:eqnx}
\gradM_{(j}V_{i)}=\mu \extK_{ij}
\end{equation}
only has solutions $(V_i,\mu)$ with $V_i=0$.  

An implicit function theorem argument \cite{BartnikFodor93} shows then
that the RTSE is solvable for all data $(\gM,J,\rho,S)$ in an
\emph{open} neighbourhood of a reference configuration
$(\mathring{\gM},\mathring{J},\mathring{\rho},\mathring{S})$
satisfying the conditions
(\ref{eq:rtsecond},\ref{eq:lapsecond},\ref{eq:eqnx}).
Here ``open'' is with respect to a Sobolev space in which the RTSE are
posed. 

The fact that the thin sandwich equations can be solved for an open
class of data $(\gM,J,\rho,S)$ can be interpreted as showing that this
reformulation of the constraints leads to ``generic'' solutions.
However, \bref{eq:rtsecond} is rather restrictive --- it excludes
asymptotically flat data for example --- so these conditions, taken
together, limit the prospects of constructing most interesting
spacetime initial data via the thin sandwich approach.

\subsection{Conformal Thin Sandwich}
\index{thin sandwich!conformal}
An interesting hybrid of the conformal method and the thin sandwich
approach has been suggested by York \cite{York99}.  Before specifying
how this {\em conformal thin sandwich} approach works, let us briefly
compare the two approaches discussed so far:
\begin{enumerate}
\item Conformal Method: The free data consists of a conformal metric
  $\lambda_{ij}$, a divergence-free trace-free tensor field
  $\sigma_{ij}$, and a scalar field $\tau$ (8 free functions). Solving
  the constraints produces a vector field $W$ and a scalar field
  $\phi$ (4 functions). We recompose to get a metric $\gM_{ij}$ and a
  symmetric tensor $K_{ij}$ satisfying the constraints (12
  functions). The lapse and shift are ignored. 
\item Thin Sandwich: The free data consists of a metric $\gM_{ij}$ and
  a symmetric tensor $J_{ij}$ (12 free functions). Solving the
  constraints produces a vector field $X^i$ and a scalar field $N$ (4
  functions).  We recompose to get a metric $\gM_{ij}$ and a
  symmetric tensor $K_{ij}$ satisfying the constraints, plus the lapse
  $N$ and the shift $X^i$ (16 functions).
\end{enumerate}

The idea of the conformal thin sandwich approach is to specify as free
data a conformal metric $\lambda_{ij}$, a trace-free symmetric tensor
$U_{ij}$, a scalar field $\tau$, and another scalar field $\eta$. In
terms of the loose  function counting system used above, this amounts
to 12 free functions. In the spirit of the thin sandwich, the tensor
$U_{ij}$ represents the time derivative of the conformal metric, while
$\tau$ is the mean curvature and $\eta$ represents the lapse
function, up to a conformal factor. It follows from this interpretation,
together with a choice of conformal scaling, that the second
fundamental form $K_{ij}$ is expressed as  
\begin{equation} 
\extK_{ij} =\psi^{-2}\tfrac{1}{2 \eta}((LX)_{ij}-U_{ij}) + \tfrac{1}{3} \psi^4 \tau \lambda_{ij}
\label{CTSK}
\end{equation}
where $X^i$ is the shift vector and $\psi$ is the conformal factor,
neither of which is known at this stage. To determine $X^i$ and
$\psi$, we use the constraint equations. These take the form 
\begin{equation}
   \label{eq:15}
   \gradM{}^j((2\eta)^{-1}(LX)_{ij}) =
   \gradM{}^j((2\eta)^{-1}U_{ij}) + \tfrac{2}{3}\psi^6 \gradM_i\tau,
\end{equation}
and
\begin{equation}
   \label{eq:16}
   8\Delta \psi - R(\gM)\psi = - A_{ij}A^{ij} \psi^{-7} + \tfrac{2}{3}
   \tau^2 \psi^5. 
\end{equation}
where we use the convenient short hand $A_{ij}=\tfrac{1}{2
\eta}((LX)_{ij}-U_{ij})$. Thus, if we can solve
(\ref{eq:15}-\ref{eq:16}) for $X$ and $\psi$ (4 functions), we have
$K$ (obtained from \bref{CTSK}) and  
\begin{equation}
\gM_{ij} = \psi^4 \lambda_{ij} 
\end{equation} 
satisfying the constraints, plus a specification of the lapse
$N=\psi^6 \eta$ and the shift $X^i$. 

Compared to the conformal method (``CM''), the conformal thin sandwich
approach (``CTSA'') has a number of virtues:
\begin{enumerate}
\item Unlike CM, specifying a set of CTSA free data does not require a
  projection to the divergence-free part of a symmetric tracefree
  2-tensor.
\item The map from CTSA free data to a solution of the constraint
  equations is conformally covariant in the sense that if
  $(\gM_{ij},\extK_{ij})$ is a solution corresponding to a particular
  choice of free data $(\lambda_{ij}, U_{ij}, \tau, \eta)$, then it is
  also a solution for the CTSA data  $(\theta^4\lambda_{ij},
  \theta^{-2}U_{ij}, \tau, \theta^6\eta)$, for any positive function
  $\theta$.  This is true whether or not $\tau$ is constant, whereas CM
  data has this property only if $\tau$ is constant.
\item The mathematical form and hence the mathematical analysis of the
  CTSA equations \bref{eq:15}--\bref{eq:16} is very similar to that of
  the CM equations \bref{confmomA}--\bref{LichneroA}.  Hence we have
  essentially the same existence and uniqueness results for the two
  sets of equations.
\item CTSA free data is arguably closer to the physics we wish to
  model, since it includes the time derivative of the conformal
  metric, while the CM data only has the divergence-free trace-free
  (transverse traceless) part of the second fundamental form.
\end{enumerate}

Along with these virtues, CTSA has one troubling feature:  Say we want
to find a set of CMC initial data $(\gM_{ij},\extK_{ij})$ with the
lapse function chosen so that the evolving data continues to have
constant mean curvature.  In the case of the conformal method, after
solving \bref{confmomA} and \bref{LichneroA} to obtain a solution of the
constraints, one achieves this by solving a linear elliptic equation
for the lapse function.  This equation is not coupled to the
constraint equations, and solutions are readily verified to exist.  By
contrast, in the CTSA the extra equation takes the form
\begin{equation}
  \label{CTSA-CMC}
\Delta(\psi^7\eta) = \tfrac{1}{8}  \psi^7\eta R
+\tfrac{5}{12}(\psi\eta)^{-1}(LX-U)^2 +\psi^5 X^i\nabla_i\tau-\psi^5,
\end{equation}
which is coupled to the CTSA equations \bref{eq:15}--\bref{eq:16}.
This coupling cannot be removed by choosing $\tau$ constant.  Indeed,
whether the data is CMC or not, the mathematical analysis of the full
system (\ref{eq:15}--\ref{CTSA-CMC}) is not very tractable.  We
emphasise, however, that if we do not require that the lapse be chosen
to preserve the mean curvature, then the analysis of the CTSA
equations is no more difficult than that of the conformal method
equations.  For further discussion of the conformal thin sandwich
approach, both in theory and in practice, see \cite{York99} and
\cite{Cook03}.

\subsection{Parabolic methods and the quasi-spherical ansatz}
\label{sec:QS}

It was first shown in \cite{Bartnik93} that
3-metrics\footnote{Although we restrict discussion here to
  3-manifolds, the parabolic method generalizes to all dimensions.} of
prescribed scalar curvature can be constructed by solving a certain
\emph{parabolic} equation on $S^2$.  This leads to a method for
constructing solutions of the Hamiltonian constraint which has found
some interesting applications
\cite{Bartnik93,SmithWeinstein00,ShiTam02}.

To describe the general construction \cite{SmithWeinstein00}, consider
the second variation identity for a foliation $r\to\Sigma_r$ in a
Riemannian 3-manifold with metric $\gQS$, \index{second
  variation!Riemannian}
\begin{equation}
   \label{2varM}
   R(\gQS)= 2 D_n H + 2K_G -H^2 - \|\II\|^2-2u^{-1}\Delta u\;,
\end{equation}
where $n$ is the outward unit normal vector to the level sets
$\Sigma_r$, $\II$ is the second fundamental form,
$H=\tr_{\Sigma_r}\II$ is the mean curvature, and $K_G$,
$\Delta=r^{-2}\Delta_o$ are respectively the Gauss curvature and
Laplacian of the metric $r^2\go$ on the foliation 2-surfaces (which
are usually assumed to be topological 2-spheres).  This is
the Riemannian version of the Lorentzian formula \bref{eq:2varV}.  The
name ``second variation'' arises from the term $D_n H$, since the mean
curvature measures the first variation of the area of $\Sigma_r$.
This interpretation is not important for the present application,
although it does reflect some deep relationship between mean curvature
and minimal surfaces on the one hand, and positivity properties of
mass on the other \cite{SchoenYau79,ShiTam02}.

Consider \bref{2varM} when the surface variation $r\to \Sigma_r$
defines a foliation, so the metric takes the general form
\begin{equation}
   \label{eq:QS}
   g = u^2\,dr^2 +  
\go_{AB}(rd\theta^A+\beta^Adr)(rd\theta^B+\beta^Bdr),
\end{equation}
where $u=u(r,\theta^A)>0$, $\beta=(\beta^A(r,\theta^B))$ and the
rescaled angular metric $\go_{AB}(r,\theta^C)$ are arbitrary fields.  The
\emph{quasi-spherical} (QS) case considered in \cite{Bartnik93} arises as
the special case where
$\go=d\vartheta^2+\sin^2\vartheta\,d\varphi^2$ is the standard
2-sphere metric.  Generally, the mean curvature of the leaves $\Sigma_r$ is 
\begin{equation}
   \label{mcSr}
   H = -\frac{1}{ru}\left(2-\divo \beta
   +r\partial_r\log\sqrt{\det\go}\right), 
\end{equation}
where $\divo\beta$ is the divergence of the tangential vector field
$\beta=\beta^A\partial_A$ in the metric $\go$, and
$n=u^{-1}(\partial_r-r^{-1} \beta^A\partial_A)$ is the
exterior-directed unit normal vector.  Assume throughout the
\emph{quasiconvexity condition}
\begin{equation}
   \label{hdef}
 h := 2-\divo\beta +r\partial_r \log\sqrt{\det \go}  \ \  >\  0,
\end{equation}
and let $n_o=un$, $H_o = uH=-h/r$, $\II_o=u\, \II$ be the corresponding
quantities determined by the metric \bref{eq:QS} with $u=1$.  The key
observation is that substituting $H$ from \bref{mcSr} into $D_nH$
gives a term $hu^{-3}r^{-1}\partial_r u$ which combines with the term
$u^{-1}\Delta u$ to show that if the scalar curvature is specified
then \bref{2varM} may be read as a parabolic equation for the lapse
$u$.  Here  $\beta^A$, $\go_{AB}$ can be freely chosen, subject only to
the quasiconvexity condition \bref{hdef}, or equivalently, $H<0$.
Explicitly we have \index{quasi-spherical equation}
\begin{eqnarray}
   \label{gQSeqn}
\lefteqn{  h (r\partial_r - \beta^A\partial_A)u\ -\  u^2 \Delta_o u}
\\
\nonumber
&= &
 ru(r\partial_r - \beta^A\partial_A)H_o + \half r^2 u(H_o^2+|\II_o|^2)
   -u^3 (K(\go)-\half R(\gQS)r^2),
\end{eqnarray}
which clearly shows the parabolic structure of the equation satisfied by $u$.

Thus if $R(\gQS)$ is determined by the Hamiltonian constraint equation
\bref{Hcon} and $\beta$, $\go_{AB}$ are regarded as prescribed fields, then
solving \bref{gQSeqn} leads to a 3-metric which satisfies the
Hamiltonian constraint.


Global ($r\to\infty$) existence theorems were established for
\bref{gQSeqn} in the quasi-spherical case
$\go=d\vartheta^2+\sin^2\vartheta\,d\varphi^2$ \cite{Bartnik93},
assuming $\beta$ satisfies the quasiconvexity condition \bref{hdef},
which becomes $\divo\beta <2$, and assuming the prescribed scalar curvature is
not too positive ($r^2R\le2$ is sufficient but not necessary).  With
suitable decay conditions on the prescribable fields $\beta^A$ (for
example, if $\beta^A$ is compactly supported), these results give
asymptotically flat solutions of the Hamiltonian constraint, with
either black hole $H=0$ or regular centre $r=0$ inner boundary
conditions.  Setting $R(\gQS)=0$ gives a large class of solutions of
the vacuum Hamiltonian constraint \bref{Hcon} in the
\emph{time-symmetric} case $\extK_{ij}=0$, with two degrees of freedom
corresponding to the choice of $\beta^A$. Analogous results have been
established for general $\go_{AB}$
\cite{SmithWeinstein00,SmithWeinstein03,ShiTam02}, which demonstrate
that this provides a flexible technique for extending solutions to be
asymptotically flat, or asymptotically hyperboloidal if $R(\gQS)=-6$.

The parabolic method enables us to construct metrics of prescribed
scalar curvature which have properties not achievable by conformal
methods.  For example, any bounded domain $(\Omega,g)$ with smooth
mean-convex ($H<0$) boundary $\partial\Omega \simeq S^2$ and
non-negative scalar curvature can be extended smoothly to an
asymptotically flat manifold, also with $R\ge0$; we simply extend the
Gaussian normal foliation in a neighbourhood of $\partial\Omega$
smoothly to a metric which is flat outside a neighbourhood of
$\Omega$, with a mean-convex exterior foliation approaching the
standard spherical foliation of $\bR^3$.  Solving the parabolic lapse
equation for $R\ge0$ with initial data $u=1$ on a Gaussian level set
$\Sigma_{-\epsilon} \subset \Omega$ produces the required 3-metric.

One consequence of this argument, first observed in \cite{Bartnik93},
is the existence of $R=0$ metrics on $\bR^3$ which are asymptotically
flat and non-flat, but which have a flat interior region.

The parabolic method also gives solutions satisfying ``geometric
boundary conditions'' \index{ geometric boundary conditions}
\cite{Bartnik97d} $(r^2\go,H)$, which specify
the boundary metric $r^2\go_{AB}$ and mean curvature $H$.  This is
clear from \bref{mcSr} since $H$ is determined by specifying the
well-posed initial condition $u(r_0)$ for the parabolic lapse
equation.  Note that this is not possible with conformal methods,
since specifying both the boundary metric and mean curvature leads to
simultaneous Dirichlet and Neumann boundary conditions on the
conformal factor.  These boundary conditions are ill-posed for the
elliptic equation imposed on the conformal factor, which will not then
admit any solution in general.  The problem of constructing 3-metrics
with prescribed boundary metric and mean curvature arises naturally
from the geometric definition of quasi-local mass \cite{Bartnik89a}
\index{quasi-local mass}
and provided the original motivation for the quasi-spherical
construction.

Examples of 3-metrics satisfying the Geroch (inverse mean curvature
flow) \index{inverse mean curvature flow} foliation condition
\cite{Geroch73,HuiskenIlmanen01} can be constructed by choosing
$\beta^A$, $\go_{AB}$ satisfying $\divo\beta =
r\partial_r(\log\det\go)$ and then solving the parabolic lapse
equation with suitably prescribed scalar curvature.  This provides a
large class of metrics for which the Geroch identity directly verifies
the Penrose conjecture. \index{Penrose conjecture} Note that this does
not help to solve the much harder problem considered in
\cite{HuiskenIlmanen01}, of finding a Geroch foliation in a
\emph{given} metric.

Another advantage of the parabolic lapse method is that it is
generally easier and cheaper numerically to solve a 2+1 parabolic
equation than the 3D elliptic equations arising in the conformal
methods, particularly when the solution is required near spatial
infinity.  Finally, in the original quasi-spherical gauge
\cite{Bartnik93}, the freely-specifiable fields $\beta^A, A=1,2$ give
a rather explicit parameterization of the ``true degrees of freedom''
\index{gravitational degrees of freedom}
allowed by the Hamiltonian constraint, since fixing $\beta$ removes
all diffeomorphism freedom in the quasi-spherical gauge (at the
linearized level at least).

\index{quasi-spherical foliations}
Dual to the problem of \emph{constructing} metrics having
quasi-spherical form, is the harder problem of \emph{finding} a
quasi-spherical foliation in a given metric.  Some simple observations
suggest that QS foliations exist for general perturbations of a QS
metric, but for interesting technical reasons a complete proof is not
yet available.  

Write the general QS 3-metric as $g = u^2dr^2+ \Sigma_A
(\beta_Adr+r\sigma_A)^2$ where $\sigma_A$, $A=1,2$ is an orthonormal coframe
for the unit sphere $S^2$, and consider the effect of an infinitesimal
diffeomorphism generated by $X=\zeta^A \tau_A + z
(r\partial_r-\beta^A\tau_A)$, where $\tau_A$ is the dual frame to
$\sigma_A$.  The QS condition will be preserved
under the metric variation $\delta g_{ij}$ and infinitesimal diffeomorphism $X$
exactly when $(\cL_Xg + \delta g)_{AB}=0$, which gives the equation
\begin{equation}
  \zeta_{(A|B)} +  (\sigma_{AB}-\beta_{(A|B)}) z  + \half
  \delta g_{AB} = 0.  
\end{equation}
Taking trace and trace-free projections and simplifying gives
\begin{eqnarray}
  \label{DB}
  \DB\zeta &:=& \zeta_{(A|B)} - \half \divo \zeta \sigma_{AB} +
  \cB_{AB} \divo \zeta 
\nonumber
\\
   &=& -\half(  \delta g_{AB} - \half \delta g_C^C \sigma_{AB} + \cB_{AB}
  \delta g_C^C),
\end{eqnarray}
where 
\[  \cB := \frac{\beta_{(A|B)} -\half \divo\beta \sigma_{AB}}{2-\divo\beta}
\]
and
\[
z = -\frac{\divo \zeta + \half \delta g^A_A}{2-\divo\beta}.
\]
The operator $\DB \zeta$ in \bref{DB} is elliptic exactly when
$|\cB|^2<1$, and we can show that its adjoint $\DB^\dagger y_{AB} =
-y_{AB}^{\ \ \ |B} + (y_{BC}\cB^{BC})_{|A}$ has trivial kernel when
$|\cB|^2 < 1/3$ (pointwise). This latter condition also guarantees
that $\DB$ is surjective, whereupon \bref{DB} can be solved for
$\zeta_A$, uniquely if the $L=1$ spherical harmonic components of
$\chi_A$ are specified.  Hence existence of QS coordinates can be
established at the linearized level, but the presence of $\divo\zeta$
in the expression for $z$ means that it is not possible to directly
apply an implicit function theorem to conclude the existence for all
metrics in an open neighbourhood of $g$.

It is also interesting to consider the direct approach to finding
quasi-spheres as graphs over $S^2$ in $\bR^3$.  From the second
fundamental form of the graph $r=e^u$, $u:S^2\to\bR$,
\[
\II_{AB} = \frac{e^u}{\sqrt{1+|Du|^2}}(u_{AB}-u_Au_B-\sigma_{AB})
\]
where $\sigma_{AB}$ is the $S^2$ metric and
$|Du|^2=\sigma^{AB}u_Au_B$,
the Gauss and mean curvatures are given by 
\begin{eqnarray}
K(u) &=& \det \II_{AB} / \det \bar{\sigma}_{AB}
\\
H(u) &=& e^{-2u} \bar{\sigma}^{AB} \II_{AB},
\end{eqnarray}
where $\bar{\sigma}_{AB}=e^{2u}(\sigma_{AB}+u_Au_B)$ is the induced
metric on the graph.  It follows that the linearisations about the
unit sphere with respect to infinitesimal changes $v=\delta u,
h=\delta g$ are
\begin{eqnarray}
  - \delta K(v,h) &=&  (\Delta +2)v + \half( (\Delta+1)
    (\sigma^{AB}h_{AB})  -  h_{AB}^{\ \ \ |AB})
\\
   H\,v &=& (\Delta +2)v + h_{rr} +\sigma^{AB}(h_{AB} +\nabla_Ah_{Br}
  -\half\partial_r h_{AB}).
\end{eqnarray}
Now the equation $(\Delta+2)\phi =f$ is solvable for $\phi$ exactly
when $\oint_{S^2} f Y_1 =0$ for any $l=1$ spherical harmonic $Y_1$, and it
follows that for generic metric variations $h$, it is not possible to
find a corresponding variation $v$ of the graph which preserves the
condition $H=-2$.  This instability of constant mean curvature
foliations can be overcome if the background metric has positive ADM
mass \cite{HuiskenYau96}.  In contrast, the equation $\delta K(v,h)=0$
is solvable for $v$ for any metric variation $h$, since
\[
\oint_{S^2}  (\Delta+1) \sigma^{AB}h_{AB}  -  h_{AB}^{|AB}) Y_1 =0
\]
for all $Y_1$ satisfying $\Delta Y_1 =-2Y_1$ and then $\nabla Y_1$ is a
conformal Killing vector of $S^2$ and $\nabla^2_{AB} Y_1 = (\Delta Y_1
+Y_1)\sigma_{AB}$.  Thus at the linearised level it is always possible
to preserve the condition $K=1$.

Further evidence supporting the conjecture that QS foliations exist
for generic metrics near a given QS metric satisfying $|\cB|^2 < 1/3$
comes from the thesis of Spillane \cite{Spillane94}, which establishes
existence in the axially symmetric case.  However, the general case
remains open, as does the related problem of finding a QS foliation of
a null hypersurface \cite{Bartnik97a}.  Note that this
\emph{foliation} existence question is irrelevant for the
\emph{metric} existence question, which is resolved by the proof in
\cite{Bartnik93} of a  large class of metrics in QS form.


The momentum constraint equations are not yet well-understood in the
parabolic method, although some results have been established
\cite{Sharples01}.  In the quasi-spherical case, we may introduce the
parameterization
\begin{equation}\label{KQS}
   \extK_{ij}\theta^i\theta^j = (\eta_{AB}+\half\mu\delta_{AB})\theta^A\theta^B
                      +\kappa_A(\theta^A\theta^3+\theta^3\theta^A) +
                      (\tau-\mu)\theta^3\theta^3 , 
\end{equation}
where $\theta^1 = \beta^1\, dr + r\,d\vartheta$, $\theta^2 =
\beta^2\,dr + r\sin\vartheta\, d\varphi$, $\theta^3 = u\, dr$ is the
QS orthonormal coframe.  In terms of the parameters $\eta_{AB}$,
$\mu$, $\kappa_A$, $\tau=\trM \extK$, the momentum constraint equations
then take the form
\begin{eqnarray}
\label{T03}
  8\pi T_{03} ru &=& -(r\partial_r\mu - \beta^A\nabla_A\mu +
        (3-\sdiv\beta)\mu) \\ 
  &&{}+ u\sdiv\kappa +\kappa^A\nabla_A u + \eta^{AB}\beta_{(A|B)}+(2-\sdiv\beta)\tau
                  \nonumber\\[3pt]
\label{T0A}
  8\pi T_{0A} ru &=&
           (u\eta_{AB})^{|B} + \mu\nabla_A u + \nabla_A(u(\half\mu-\tau))   \\
    &&{}+r\partial_r \kappa_A -\beta^B\kappa_{A|B}
           +((3-\sdiv\beta)\delta_{AB}-\beta_{B|A})\kappa^B. 
                       \nonumber
\end{eqnarray}
The matter fields $T_{00}, T_{0A}, T_{03} $ are prescribed and
\bref{T03} provides either an equation for $\tau$, by choosing $\mu$,
or vice-versa, choosing $\tau$ and solving for $\mu$. Likewise,
\bref{T0A} can be regarded either as an equation for $\kappa$ (with
the symmetric traceless 2-tensor $\eta_{AB}$ arbitrarily prescribable)
or as an elliptic equation for $\eta$ with $\kappa_A$ freely
prescribable.

Local existence in $r$ for the momentum constraints with prescribed
$\tau,\kappa_A$ has been  established by Sharples \cite{Sharples01}, but it
is not clear whether global results are possible without additional
restrictions.  Much work still remains to be done on these systems.

\index{null quasi-spherical coordinates}
Finally we note that a characteristic version of the quasi-spherical
gauge, where a foliation of outgoing ($\rho_{NP}<0$) null
hypersurfaces is assumed to admit a QS radial coordinate, has been 
described in \cite{Bartnik97a}.  In this case there is no parabolic
equation.  The resulting hypersurface Einstein equations are considerably
simpler than those derived in the Bondi gauge \cite{Sachs62,Bondi60},
and forms the basis for a 4th order numerical code
\cite{BartnikNorton00}, which heavily exploits the exact spherical
geometry of the $r$-level surfaces.

\subsection{Gluing Solutions of the Constraint Equations}

The conformal method, the conformal thin sandwich method, and the
quasi-spherical ansatz are all procedures for generating solutions of
the Einstein constraint equations from scratch. We now 
consider procedures for constructing new solutions of the
constraints from existing solutions.  

\index{gluing constraint solutions}
We first discuss a procedure for gluing connected sums of solutions.
The idea of this ``IMP Gluing'' \cite{IsenbergMazzeoPollack02} is the
following: Say we have two solutions of the constraint equations,
$(M_1, \gamma_1, K_1)$ and $(M_2, \gamma_2, K_2)$. Let $p_1\in M_1$
and $p_2 \in M_2$. Can we find a set of initial data $(M_{(1-2)},
\gamma_{(1-2)}, K_{(1-2)} )$ such that 1) $M_{1-2}$ is homotopic to
the connected sum\footnote{The connected sum of these two manifolds is
  constructed as follows: First we remove a ball from each of the
  manifolds $M_1$ and $M_2$. We then use a cylindrical bridge $S^2
  \times I$ (where $I$ is an interval in $R^1$) to connect the
  resulting $S^2$ boundaries on each manifold.}  $M_1\#M_2$; 2)
$(\gamma_{(1-2)}, K_{(1-2)})$ is a solution of the constraints
everywhere on $M_{(1-2)}$; and 3) on that portion of $M_{(1-2)}$ which
corresponds to\\ $M_1\setminus\{ \textrm{ball around } p_1\}$, the
data $(\gamma_{(1-2)}, K_{(1-2)})$ is isomorphic to $(\gamma_1, K_1)$,
with a corresponding property holding on that portion of $M_{(1-2)}$
which corresponds to $M_2\setminus\{\textrm{ball around }p_2\}$?  If
so, we say that the sets of data admit IMP gluing.

IMP gluing can be carried out for quite general sets of initial data.
The sets can be asymptotically Euclidean, asymptotically hyperbolic,
specified on a closed manifold, or indeed anything else. The only
condition the data sets must satisfy is that, in sufficiently small
neighborhoods of each of the points at which the gluing is to be done,
there do not exist nontrivial solutions $\xi$ to the equation
$D\Phi^{*}_{(\gamma, K)} \xi=0$, where $D\Phi^{*}_{(\gamma, K)} $ is
the linearization operator defined in \bref{DPhi*} (with $K$ replacing
$\pi$). In \cite{BeigChruscielSchoen04} it is shown that this
condition is indeed generic.


The proof that IMP gluing can be carried out to this degree of
generality is detailed in \cite{ChruscielIsenbergPollack04}, based on
\cite{IsenbergMazzeoPollack02}, \cite{IsenbergMazzeoPollack03}, and
\cite{ChruscielDelay02}. We note here two features of it.  First, the
proof is constructive, in the sense that it outlines a systematic,
step-by-step mathematical procedure for doing the gluing: One
conformally blows up the balls surrounding $p_1$ and $p_2$ to produce
two half cylinders extending from the original initial data sets; one
joins the two half cylinders into a bridge, and splices together the
data from each side using cutoff functions; one uses the local
constant mean curvature to decouple the constraints in the
neighborhood of the bridge; one uses tensor projection operators based
on linear PDE solutions to find a new conformal $K$ which solves the
momentum constraint; one solves the Lichnerowicz equation
\index{Lichnerowicz equation} (the
argument that this can be done, and that the solution is very close to
1 away from the bridge, relies on the invertiblity of the linearized
equation and on a contraction mapping); one recomposes the data as in
\bref{recongammaA} and \bref{reconKA}; and finally one does a
nonconformal data perturbation away from the bridge to return the data
there to what it was before the gluing. This procedure can be largely
carried out numerically, although we note that it requires us to solve
elliptic equations on topologically nontrivial manifolds.

The second feature we note regarding the proof is that it  relies
primarily on the conformal method, but it also uses a nonconformal
deformation of the data at the end, to guarantee that the glued data
is not just very close to the given data on regions away from the
bridge, but is exactly equal to it. 

While IMP gluing is not the most efficient tool for studying  the
complete set of solutions of the constraints, it has already proven to
be very useful for a number of applications, including the
following:
\begin{enumerate}
\item {\it Multi-Black Hole Data Sets:} Given an asymptotically
  Euclidean solution of the constraints, IMP gluing allows a sequence
  of flat space initial data sets to be glued to it. The bridges that
  result from this gluing each contain a minimal surface, and
  consequently an apparent horizon. With a bit of care
  \cite{ChruscielDelay02}, one can do this in such a way that indeed
  the apparent horizons are disjoint, and therefore likely to lead to
  independent black holes.
\item {\it Adding a Black Hole to a Cosmological Spacetime:} Although
  there is no clear established definition for a black hole in a
  spatially compact solution of Einstein's equations, one can glue an
  asymptotically Euclidean solution of the constraints to a solution
  on a compact manifold, in such a way that there is an apparent
  horizon on the bridge. Studying the nature of these solutions of the
  constraints, and their evolution, could be useful in trying to
  understand what one might mean by a black hole in a cosmological
  spacetime.
  
\item {\it Adding a Wormhole to Your Spacetime:}
  \index{wormhole!construction}  While we have
  discussed IMP gluing as a procedure which builds solutions of the
  constraints with a bridge connecting two points on different
  manifolds, it can also be used to build a solution with a bridge
  connecting a pair of points on the {\it same} manifold. This allows
  one to do the following: If one has a globally hyperbolic spacetime
  solution of Einstein's equations, one can choose a Cauchy surface
  for that solution, choose a pair of points on that Cauchy surface,
  and glue the solution to itself via a bridge from one of these points
  to the other. If one now evolves this glued-together initial data
  into a spacetime, it will likely become singular very quickly
  because of the collapse of the bridge. Until the singularity
  develops, however, the solution is essentially as it was before the
  gluing, with the addition of an effective wormhole. Hence, this
  procedure can be used to glue a
  wormhole onto a generic spacetime solution. 
\item {\it Removing Topological Obstructions for Constraint
    Solutions:} We know that every closed three dimensional manifold
  $M^3$ admits a solution of the vacuum constraint equations. To show
  this, we use the fact that $M^3$ always admits a metric $\Gamma$ of
  constant negative scalar curvature. One easily verifies that the
  data $(\gM=\Gamma, K=\Gamma)$ is a CMC solution. Combining this
  result with IMP gluing, one can show that for every closed $M^3$,
  the manifold $M^3 \setminus \{p\}$ admits both an asymptotically
  Euclidean and an asymptotically hyperbolic solution of the vacuum
  constraint equations.
  
\item{\it Proving the Existence of Vacuum Solutions on Closed
    Manifolds with No CMC Cauchy Surface:} \index{constant mean
    curvature!nonexistence}
  Based on the work of
  Bartnik \cite{Bartnik88a,Bartnik88b}, one can show that if one has a set of
  initial data on the manifold $T^3\#T^3$ with the metric components
  even across a central sphere and the components of $K$ odd across
  that same central sphere, then the spacetime development of that
  data does not admit a CMC Cauchy surface. Using IMP gluing, one can
  show that indeed initial data sets of this sort exist.

\end{enumerate}

\subsection{The Corvino-Schoen method}

There is another very useful form of gluing which has been applied
recently to construct interesting solutions of the Einstein vacuum
constraint equations.  Developed by Corvino and Schoen
\cite{Corvino00,CorvinoSchoen03,ChruscielDelay02}, this method has the
following remarkable application.  Let $(M^3, \gM, \extK)$ be a smooth,
asymptotically Euclidean solution of the constraint equations. If
certain asymptotic conditions hold, then for any compact region
$\Sigma^3 \subset M^3$ for which $M^3 \setminus \Sigma^3 =
\mathbb{R}^3 \setminus B^3$ (where $B^3$ is a ball in $\mathbb{R}^3$),
there is a smooth asymptotically Euclidean solution on $M^3$ which
is identical to the original solution on $\Sigma^3 \subset M^3$, and
is identical to Cauchy data for the Kerr solution on $M^3 \setminus
{\tilde \Sigma^3}$ for some ${\tilde \Sigma^3} \subset M^3$. In words,
their technique allows us to smoothly glue any interior region of an
asymptotically Euclidean solution to an exterior region of a slice of
the Kerr solution. For asymptotically Euclidean solutions of the
constraints with $\trM \extK=0$, this method glues any interior region to
an exterior region of a slice of Schwarzschild.


Combining the Corvino-Schoen gluing techniques with some results of
Friedrich \cite{Friedrich85}, \cite{ChruscielDelay02} showed that
there is a large class of vacuum spacetime solutions of Einstein's
equations which admit complete null infinity regions of the form
``scri'', as hypothesized by Penrose. The tools developed by Corvino
and Schoen have also been used to strengthen the IMP gluing results
\cite{ChruscielDelay03}.  \index{scri} \index{asymptotically simple
  spacetimes}


The Corvino-Schoen method aims to solve the constraint equations
through a projection using the linearized operator $D\Phi$
\index{linearized constraint operator} and its
adjoint $D\Phi^*$.  We sketch the method in the time-symmetric case
$\pi=0$, where $\Phi(\gM,\pi)$ is replaced by the scalar curvature
$R(\gM)$ and the lapse-shift $\xi$ is replaced by the lapse $N$. In
this case the arguments are essentially the same while the
calculations are considerably simpler.

We start with the observation that because $DR^*$ has injective
symbol, it satisfies an elliptic estimate on any domain $\Omega$,
\begin{equation}
\label{DF*1}
      \|N\|_{H^2(\Omega)} \le C ( \|DR^*N\|_{L^2(\Omega)} 
                  + \|N\|_{L^2(\Omega)} ) ,
\end{equation}
which importantly does not require any control on $N$ at the
boundary $\partial\Omega$.  It follows easily that similar weighted
estimates hold, with weight function $\rho\in C^\infty_c$ which is
positive in $\Omega$ and vanishes to high order at $\partial \Omega$:
\begin{equation}
\label{DF*2}
\|N\|^2_{H^2_\rho(\Omega)} :=  \int_\Omega \rho (N^2+|\nabla^2N|^2) \,dv_\gM
\le C 
  \int_\Omega \rho |DR^*N|^2 + N^2 )\,dv_\gM ,
\end{equation}
where the final term $N^2$ on the right can be removed if there are no
Killing vectors. \index{Killing initial data} With this assumption,
for all $f\in L^2_\rho(\Omega)$ we can solve $DR(\rho DR^*N) = \rho f$
for $N \in H^4_{\mathrm{loc}}$, which in particular produces a
solution to the linearized constraint equation $DR \,h =\rho f$.  An
iteration argument is used to solve the nonlinear problem $R(\gM_0+h)
= R(\gM_0) + S$ for any sufficiently small $S$.  This solution $h\in
H^2_{\rho^{-1}}(\Omega)$ has the remarkable property that it vanishes
to high order on $\partial \Omega$.  Thus, for example, if $R(\gM_0)$
is sufficiently small and supported in $\Omega$ then there is a
perturbation $h$, also supported in $\Omega$, such that
$R(\gM_0+h)=0$.

To use this method to glue a Schwarzschild exterior to an
asymptotically flat $R(\gM)=0$ metric across an annulus
$B_{2R}\backslash B_R$, $R >> 1$, requires one more idea because the
flat space kernel $\ker DR^*_\delta=\mathrm{span}(1,x^1,x^2,x^3)$ is
non-trivial.  This implies that the linearized problem $DR_\delta h
=\sigma$ is solvable if and only if $\sigma$ satisfies the four
conditions $\int_\Omega \sigma(1,x^i) d^3x=0$, and the nonlinear
problem $R(\gM_0+h)=0$ is similarly obstructed for $\gM_0$ close to
flat.  By choosing the cutoff radius $R$ sufficiently large and
rescaling back to $\Omega =B_2 \backslash B_1$ produces exactly this
close to flat situation.  However, it is possible to solve the
projected problem $R(\gM_0+h) \in K := \mathrm{span}(1,x^i)$ with
uniform estimates on $h\in H^2_{\rho^{-1}}(\Omega)$.  Now the
Schwarzschild exterior metric can be characterised by the mass and
centre of mass parameters $(m,c^i)$, defined by
\begin{equation}
  \gM_{\mathrm{Schw}} = (1+
  \tfrac{m}{2|\mathbf{x}-\mathbf{c}|})^4\delta_{ij}.
\end{equation}
Some delicate estimates show that the map $(m,c^i) \mapsto K$ is
continuous and has index $1$, so there is a choice of parameters
$(m,c^i)$ mapping to $0\in K$, which gives $R(\gM_0+h)=0$ as required.
The extension of these ideas, and the considerable details of the
above arguments, are given in the original references
\cite{Corvino00,CorvinoSchoen03,ChruscielDelay03}.


%

\section{Conclusion}

A considerable amount is known concerning the solutions of the
Einstein constraint equations. Using the conformal method or the
conformal thin sandwich method, we know how to construct constant mean
curvature solutions which are asymptotically Euclidean, asymptotically
hyperbolic, or live on a closed manifold. We also know how to do the
same for nearly constant mean curvature solutions. We can glue
together quite  general solutions of the constraints, producing new solutions 
of both mathematical and physical interest. And, for certain
asymptotically Euclidean solutions, we know how to show that there are
solutions which include any compact region of the solution in the
interior, and which are exactly Kerr or Schwarzschild in the exterior.

Much remains to discover as well. We would like to know how to
construct solutions with mean curvature neither constant nor nearly
constant. We would like to know much more about constructing solutions
of the constraints with prescribed boundary conditions. We would like
to know to what extent we can construct solutions with low
regularity. And we would like to know which solutions on compact
regions can be smoothly extended to either asymptotically Euclidean or
asymptotically hyperbolic solutions.  

Besides these mathematical issues to resolve, there are important
questions concerning solutions of the constraints and physical
modeling. In view of the pressing need to model astrophysical events
which produce detectable  amounts of gravitational radiation, one of
the crucial questions we need to answer is how to systematically find
solutions of the constraint equations which serve as physically
realistic initial data sets for such astrophysical models. Since these
models are generally constructed numerically, an equally crucial
question is the extent to which the constraint functions, initially
zero, remain near zero as the spacetime is numerically evolved.  

It is not clear how close we are to resolving these mathematical and
physical questions regarding the Einstein constraint equations and
their solutions. However, in view of the recent rapid progress that
has been made in these studies, we are optimistic that many of them
will be resolved soon as well. 

\section{Acknowledgments}

We thank Yvonne Choquet-Bruhat, Rafe Mazzeo, Vincent Moncrief, Niall
O'Murchadha, Daniel Pollack and James York for useful
conversations. JI thanks the Caltech relativity group and the Kavli
Institute for Theoretical Physics for hospitality  while this survey
article was being written, and acknowledges  support for this research
from NSF grant PHY 0099373 at the University of Oregon. The work of RB
is supported in part by the Australian Research Council.


\def\cprime{$'$}
\providecommand{\bysame}{\leavevmode\hbox to3em{\hrulefill}\thinspace}

\end{document}